 \documentclass[preprint2]{aastex}

\defcitealias{Pereyra:2004}{PM04}
\usepackage{url}
\usepackage{amsmath}

\slugcomment{To appear in Publications of the Astronomical Society of the Pacific.}

\shorttitle{Pipeline for polarimetric data}
\shortauthors{Ram\'irez et al.}

\begin{document}


\title{{\sc Solvepol}: a reduction pipeline for imaging polarimetry data}


\author{Edgar A. Ram\'irez}
\affil{Instituto de Astronomia, Geof\'isica e Ci\^encias Atmosf\'ericas, Universidade de S\~ao Paulo (IAG-USP), SP 05508-900, Brazil.}
\affil{Instituto Nacional de Astrof\'isica, Optica y Electr\'onica (INAOE), 72000 Puebla, Mexico.}
\email{e.ramirez@usp.br, e.ramirez@inaoep.mx}
\author{Ant\^onio M. Magalh\~aes}
\affil{Instituto de Astronomia, Geof\'isica e Ci\^encias Atmosf\'ericas, Universidade de S\~ao Paulo (IAG-USP), SP 05508-900, Brazil.}
\author{James W. Davidson Jr.}
\affil{Astronomy Department, University of Virginia, 530 McCormick Rd. Charlottesville, VA 22904-4325, USA.}
\author{Antonio Pereyra}
\affil{Instituto Geof\'isico del Per\'u, \'Area Astronom\'ia, Calle Badajoz 169, Lima, Per\'u.}
\and
\author{Marcelo Rubinho}
\affil{Instituto de Astronomia, Geof\'isica e Ci\^encias Atmosf\'ericas, Universidade de S\~ao Paulo, SP 05508-900, Brazil.}





\begin{abstract}

We present a newly, fully automated, data pipeline: {\sc solvepol}, designed to reduce and analyze polarimetric data. It has been optimized for imaging data from the Instituto de Astronom\'ia,  Geof\'isica e Ci\^encias Atmosf\'ericas (IAG) of the University of S\~ao Paulo (USP), calcite Savart prism plate-based IAGPOL polarimeter. {\sc Solvepol} is also the basis of a reduction pipeline for the wide-field optical polarimeter that will execute SOUTH POL, a survey of the polarized Southern sky. {\sc Solvepol} was written using the  {\sc interactive data language } ({\sc idl}) and is based on the {\sc image reduction and analysis facility} ({\sc  iraf}) task {\sc pccdpack}, developed by our polarimetry group. We present and discuss reduced data from standard stars and other fields and compare these results with those obtained in the {\sc  iraf} environment. Our analysis shows that {\sc solvepol}, in addition to being a fully automated pipeline, produces results consistent with those reduced by  {\sc pccdpack} and reported in the literature.

\end{abstract}


\keywords{Physical data and processes: polarization. Interstellar Medium (ISM), Nebulae: ISM: general. Astronomical Instrumentation, Methods and Techniques: techniques: image processing.}

\newpage


\section{Introduction}

The observational basis for astrophysics has been generally developed by measuring the radiation flux or brightness of objects, either in broad-band or in spectral form, regardless of the polarization state of the light beam. The polarization of the light has encoded information that can be used to investigate many astronomical phenomena in more detail, such the interstellar medium (ISM), the cosmic microwave background (CMB), active galactic nuclei (AGN), magnetic cataclysmic variables (polars), among others.



In the last few decades, a grater number of polarimetry and spectropolarimetry studies have been done \citep[eg.,][]{Trujillo-Bueno:2002,  Adamson:2005, Bastien:2011}. However,  large area polarimetric surveys are few, even though they provide an increased database for a much better knowledge of the aforementioned areas of the astronomy.


Efforts towards improving this have led to several surveys in the past few years. Near-infrared (NIR) Galactic Plane Infrared Polarization Survey \citep[GPIPS; ][]{Clemens:2012}, is a notable one. Another example is the  Southern Interstellar Polarization survey \citep{Magalhaes:2005}, aimed at obtaining optical polarization towards selected areas of (mostly) the Galaxy, in and out of the Plane. SOUTH POL \citep{Magalhaes:2012}, a broad band, all-sky polarimetric survey in the optical is slated to begin in 2017.

Surveys and/or large area programs, by their nature, typically need to have some form of data reduction pipeline. A 2k$\times$2k charge-coupled device (CCD) on a moderate, meter-sized telescope might provide several hundred polarized stars from a $\sim \! 10' \times 10'$ field. SOUTH POL and its 9k$\times$9k CCD will image  a $\sim \! 2$~degrees$^2$ field of view, roughly two orders of magnitude larger, producing tens of thousand of stars per field.  With tens of fields observed each night on a shared dedicated telescope, the need to go from the raw frames to a catalog  containing photometric, polarimetric and astrometric data in an efficient way becomes thus evident.

Over the years we have developed {\sc pccdpack} \citep{Pereyra:2000,Pereyra:2004},  an  {\sc image reduction and analysis facility} ({\sc iraf}) package for reducing polarimetric data. {\sc Pccdpack} is used at certain astronomy centers in Brazil, Chile, USA and elsewhere.  SOUTH POL and its $\sim 2$~degrees$^2$ field of view, along with a shared, dedicated robotic telescope, presented a data reduction challenge given the much large data sets. This paper presents  {\sc solvepol}, a pipeline to address this need.  The main advantage of {\sc solvepol} over the previous {\sc pccdpack} routines is that the pipeline reduces the data fully automatically without user interaction -- the previous reduction package is semiautomatic -- , speeding up the reduction process.


This paper is organized as follows. The description of the data is presented in section \ref{data:acquisition}. The description of the pipeline is given in section \ref{description:pipeline}. Results of pilot studies on polarized standard stars, the field  HD110984 and the Musca Dark Cloud, are presented in section \ref{results}. A comparison of the photometry performed by {\sc idl} and {\sc iraf} is given in section \ref{photo:section}. The Conclusion presents final considerations.

\section{Data acquisition}\label{data:acquisition}

The polarimetric data that we have analyzed with this new pipeline was acquired with the IAG-USP optical/NIR imaging polarimeter, IAGPOL \citep{Magalhaes:1996}, mounted on the 60-cm  or 1.60-m telescopes at Observat\'orio do Pico dos Dias (OPD). IAGPOL consists of a half-wave plate that can be rotated in steps of $22.5$~degrees, a calcite Savart prism, and the CCD detector.   To get the polarimetric properties of an object, a sequence of 4, 8, 12 until 16 maximum observations at adjacent positions of the waveplate are required.  For example, a bright object can be observed  with a sequence of one exposure per wavelength  position, lets say 4 positions (0, 22.5, 45 and 67.2 degrees); while a dim object, multiple exposures  per waveplate position would be more appropriate. The more positions of the waveplate, the more accurate is the polarimetry. The wide area imaging polarimeter for the SOUTH POL survey operates with exactly the same principles, other than the larger size of the optical components.



The calcite Savart plate used in IAGPOL consists of two identical prisms, each cut with their optical axis at 45 degrees to the entrance (and exit) faces, maximizing the beam separation. The two prisms were then cemented together with their optical axis crossed. If a single prism is used, the extraordinary (e) beam will suffer from astigmatism and color dispersion. By cementing the two identical prisms, the ordinary (o) ray coming from the first prism becomes the extraordinary ray for the second prism, and vice-versa, canceling to a good degree the effects present with a single prism. Also, the o-e and e-o beams will focus essentially on the same plane. 

\begin{figure*}
\centering{
\includegraphics[height=7cm]{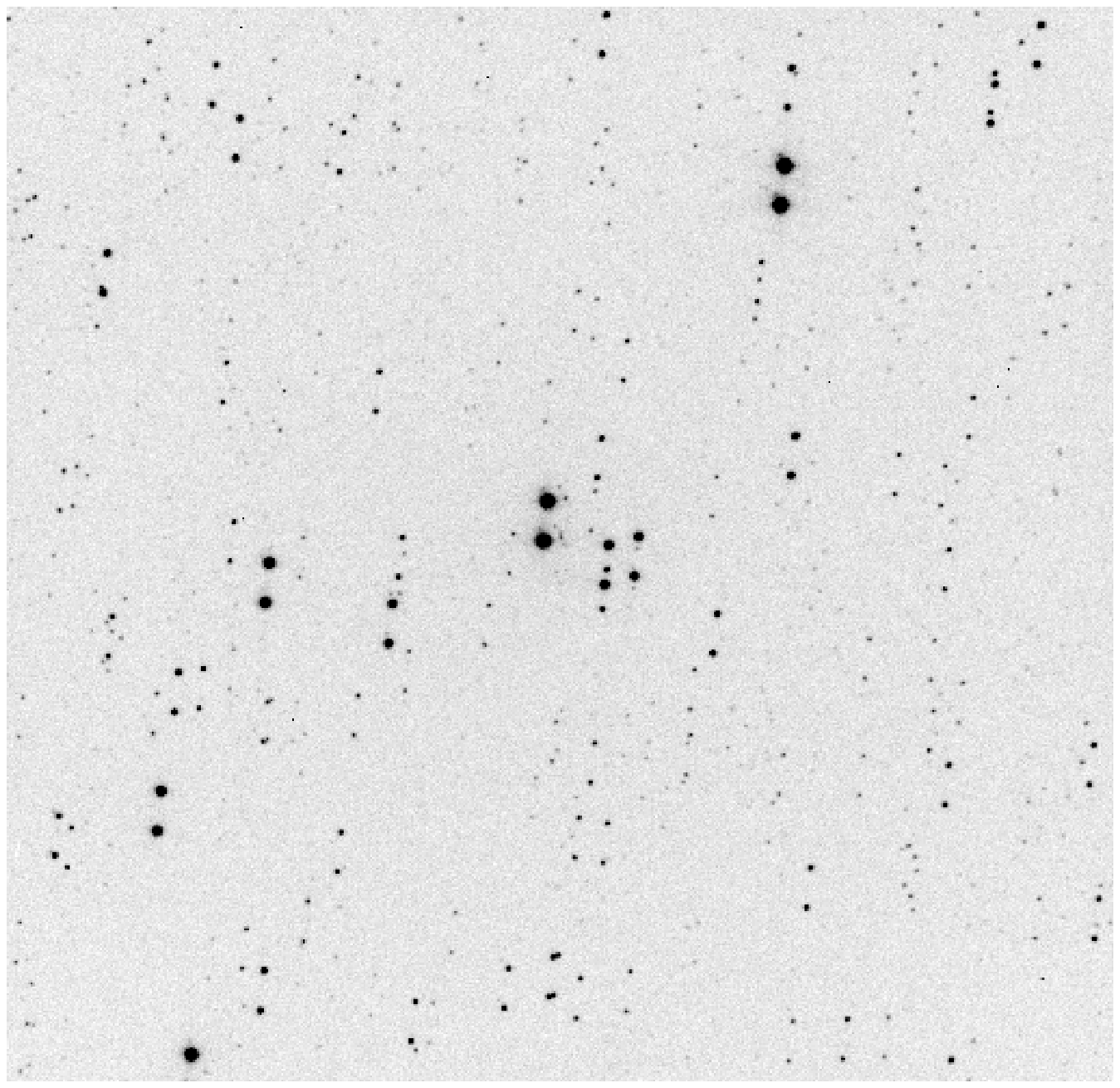} 
\includegraphics[height=7cm]{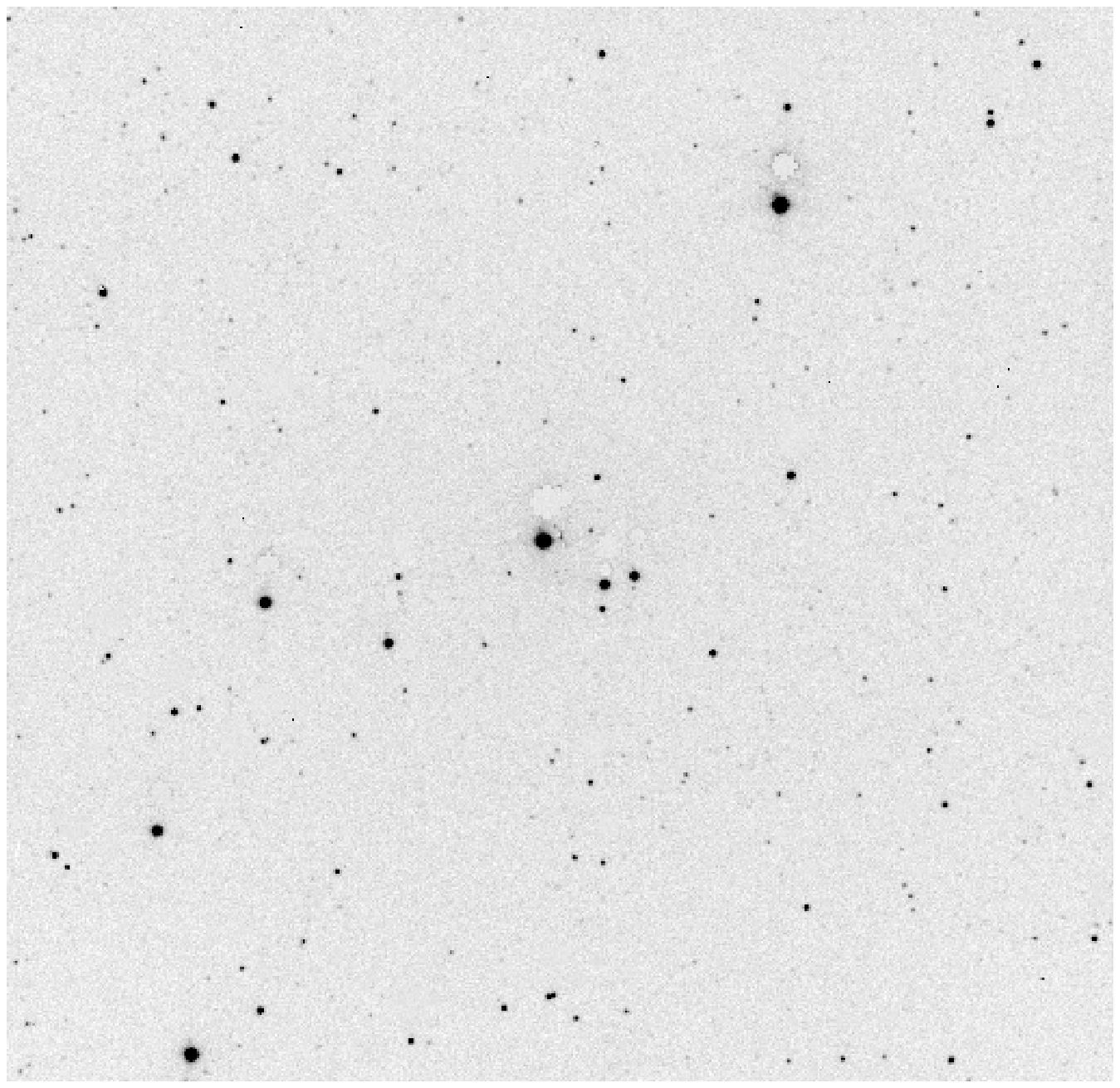}
\caption{Observations with IAGPOL \citep{Magalhaes:1996} mounted on the 0.6-m telescope at OPD, Brazil. Left: Field centered on HD110984 ($18\times 18$ arcmin) showing a typical image with the doublet of each star (ordinary and extraordinary) produced by the calcite Savart prism. Right: Same field masked out by {\sc solvepol} to calculate the astrometry (see section \ref{astrometry:calibration}). 
}
\label{FieldDoublet}
}
\end{figure*}

The calcite Savart prism causes each star image to become  double on the CCD. Figure \ref{FieldDoublet} shows a typical image obtained when using the IAGPOL showing the star doublets.  An advantage of the calcite Savart prism is that, because the ordinary and extraordinary images are perpendicularly polarized to each other, the polarization of the sky around the o-e stellar image is cancelled out \citep[eg.,][]{Magalhaes:1996}, allowing observations under non-photometric conditions. All the data analyzed has been obtained in the $V$ filter. During a typical observational campaign a minimum of one polarized standard star and one non-polarized standard star are observed per night. Several calibrators are thus  obtained during each observing run. The unpolarized standard stars are used to obtain the telescope's instrumental polarization, to be subtracted from the measured polarization. The polarized standards are used to transform the instrumental polarization position angle to one in the equatorial system.

\section{Description of the pipeline}\label{description:pipeline}

{\sc Solvepol} is written in {\sc interactive data language} ({\sc idl}) and it uses functions and procedures that are  found in the {\sc idl} Astronomy Library \citep{Landsman:1993} and in the Coyote Graphics Library\footnote[1]{\url{http://www.idlcoyote.com/}}. It also makes use of {\sc astrometry.net} suite \citep{Lang:2010}\footnote[2]{{\sc Astrometry.net} is a project partially supported by the US National Science Foundation, the US National Aeronautics and Space Administration, and the Canadian National Science and Engineering Research Council. For more information go to \url{http://astrometry.net}.}, a  very useful software for astrometric purposes that is called by the pipeline. 

{\sc Solvepol} corrects the raw images for bias and flats, and calculates the polarization ($P$), polarization position angle ($\theta$) and the catalog calibrated magnitude ($V$ magnitude, in this case) of the detected stars in the field. The main final product is a catalog in plain text format which also contains the right ascension (RA) and declination (Dec) of all stars. The polarization signal-to-noise, $P/\sigma_P$, and the ratio $F/\sigma_s$, where $F$ is the flux  of a star and $\sigma_s$ is the standard deviation of sky brightness, are initial parameters that can be set by the user and are used to reject weak sources. The complete list of input parameters is described in the manual of {\sc solvepol} available at  \url{http://www.astro.iag.usp.br/~ramirez/pipeline.html}.

In addition, we developed two procedures to help analyze the data generated by {\sc sovepol}: {\sc merge} and {\sc filter}.  

\begin{itemize} 
\item {\sc Merge} merges two catalogs calculating the weighted mean of $P$, $\theta$ (in $Q-U$ space) and $V$ of the stars common to both catalogs. {\sc Merge} reads the position, RA and Dec, of stars in two catalogs, and finds  stars that match within a  box with $0.001$ degrees (3.6 arcsec) side. 

\item {\sc Filter} extracts a subset of a catalog created by {\sc solvepol} or {\sc merge} by setting lower and upper limits to $P$, $\theta$, $V$, and lower limits to $P/\sigma_P$ and $F/\sigma_s$. These limits are set by the user in interactive mode for an ad hoc analysis. 
\end{itemize}


\subsection{CCD frame calibration}

The data reduction procedure that the pipeline applies is based on the standard procedure explained by \citet{Massey:1997}. Roughly the procedure is as follows:

\begin{itemize}

\item First, the bias frames are median combined. Then, using the {\sc idl sigma\_filter} procedure, sigma clipping is applied to remove pixels that deviate by more than  2.5 sigma from the median value of the neighboring 300 by 300 pixels. Finally, the overscan region is fit with a  polynomial of order 2 and subtracted to create the master-bias frame.

\item For flat field frames, the overscan region is fit as described previously and is subtracted to each flat field frame. Next, the flat fields are median combined. Then, sigma clipping is applied as before, and the master-bias is subtracted to create the master-flat frame. 
 
\item Each source image is reduced by first fitting a polynomial of 2nd order to the overscan region, and subtracting this fit from the image. The master-bias is then subtracted and the result is then divided by the master-flat frame normalized to the median of the sky. Finally, the source images are trimmed to remove the overscan region.

\end{itemize}

If multiple images were observed per position angle of the waveplate, they are median combined after performing the reduction steps outlined above. The pipeline estimates and corrects for any possible shifts in  $X$ and $Y$ between images before combining.

\subsection{Polarization data reduction}

The pipeline finds all point sources, i.e., stars, and performs photometric measurements which are used to calculate their polarization.  {\sc solvepol} is not going to find extended sources, and they  will be treated as background. Aperture photometry is performed on both ordinary and extraordinary images of each star  ($F_o$ and $F_e$ respectively) with ten concentric apertures ranging in size from 1 to 10 pixels in radius. For each aperture an annulus, with fixed inner and outer radii of 10 and 20 pixels respectively, is used to estimate the sky (default annulus in {\sc solvepol}). 

\begin{figure}[t]
\centering{
\includegraphics[width=6cm]{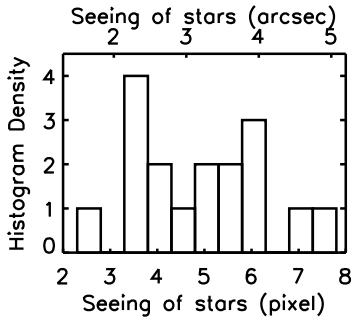}
}
\caption{Seeing of standard stars observed with the 60-cm telescope of the OPD in pixel and arcsec units. The seeing ranges from 2 to 8 pixels, thus the 10 pixels concentric apertures to perform the polarimetric measurements is justified.}
\label{fwhm:plot}
\end{figure}

These apertures cover  up to $\sim 5\times$ the full width at half maximum (FWHM) that the stelar disc may have when observing with the IAGPOL at the OPD.  Figure \ref{fwhm:plot} shows the seeing measured in standard stars observed with the 60-cm telescope. The seeing that we get ranges from 2 to 8 pixels (1.3 arcsec to 5.2 arcsec), but we know that the seeing of stars distorted due to cirrus is 7 to 8 pixels (4.5 arcsec to 5.2 arcsec). Under good observational conditions, we measured a seeing of $\sim3.5$ pixels (2.3 arcsec). Thus, the 10 concentric apertures will cover all seeing conditions.

The photometry is measured using the same aperture for all waveplate positions, and the aperture size that gives the lowest error in the polarimetry is selected to calculate the final polarimetric properties of a star. 

The modulation of the intensity, $z_i$,  where $i$ is one of sixteen orientations of the half-wave plate (i.e., $i=1$ to $4, 6, 8$ or $16$ positions), is

\begin{equation}\label{zequation}
\begin{split}
& \frac{F_{e,i} -  F_{o,i}(F^T_e/F^T_o)}{F_{e,i}+F_{o,i}(F^T_e/F^T_o)} \\
&= Q \, cos (4\,\psi_i) + U \, sin (4\,\psi_i),
\end{split}
\end{equation}

\noindent where $F_{o,i}$ and $F_{e,i}$ are the fluxes of the ordinary and extraordinary images at the position $i$ of the waveplate,  and $F^T_o$ and $F^T_e$ are the ordinary and extraordinary total flux of a star summed on all the waveplate's positions, respectively \citep[see][]{Magalhaes:1984}.   $Q$ and $U$ refer to the instrumental Stokes parameters of the incoming beam's linear polarization. Equation \ref{zequation} shows that a set of four waveplate positions (e.g., $i = 1$ through $4$) measures $Q$, $U$, $-Q$ and $-U$. A linear polarization measurement can then be obtained by doing exposures through 4, 8, 12 or 16 positions of the waveplate, although this cycle can be repeated as needed. The polarimetric properties are estimated following the  formulation of \citet{Magalhaes:1984},  which consists in solving for $U$ and $Q$ the expression \label{zequation} applying the method of the least squares.
 

The solution for the stokes parameters $Q$ and $U$ for each source are: 

\begin{equation}\label{Qsol}
Q=  \frac{2}{\mu} \sum_i^{\mu} z_i cos (4\psi_i)
\end{equation}
\begin{equation}\label{Qsol}
U= \frac{2}{\mu} \sum_i^{\mu} z_i sin (4\psi_i),
\end{equation}

\noindent where $\mu$ is the number of positions of the half-wave plate and  $\psi_i=[i-1]*22.5$~degrees ($=0, 22.5, 45, 67.5$ degrees, etc.). This yields the percent linear polarization, $P$, and  the polarization angle, $\theta$, measured from north to east after the zero-angle is calibrated using polarized standard stars: 

\begin{equation}
P=\sqrt{Q^2 + U^2}
\end{equation}
\begin{equation}
\theta = \frac{1}{2} tan^{-1} \frac{U}{Q}.
\end{equation}


The uncertainties of the polarimetric properties are in essence the residual of the actual measurements at each waveplate position angle with respect to the theoretical non-linear curve of the modulation. Assuming that $ \sigma_Q  = \sigma_U  = \sigma_P$, the error may be calculated as \cite[see ][]{Magalhaes:1984,Naghizadeh-Khouei:1993}

\begin{equation}
\sigma_P =  \frac{1}{\sqrt {   \mu-2}}   \sqrt { \frac{2}{\mu}  \sum_i^{\mu} z_i^2 - Q^2 - U^2}
\end{equation}
\begin{equation}
\sigma_{\theta} = 28^{\circ}.65 \frac{\sigma_P}{P}.
\end{equation}

\begin{figure}[t]
\centering{
\includegraphics[width=6cm]{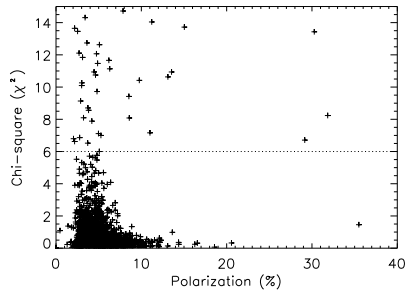}
}
\caption{Chi square of the modulation fit vs Polarization of the stars in Musca fields analyzed in this work.  $\chi^2 > 6$ is a reasonable limit to reject stars with bad modulation fit and therefore spurious polarization. The selection is $F/\sigma_s\geq 5.0$ and $P/\sigma_P\ge 5.0$. Notice that there is a spurious star with $\chi^2  \approx 1 $ and $P\approx 35$ that the user should distinguish.}
\label{chisqrt}
\end{figure}

The reduced Chi square ($\chi^2$) of the fitted modulation is calculated, and for this work, those stars with $\chi^2 > 6.0$ are rejected.  This limit was set by empirical grounds. Figure \ref{chisqrt} shows the $\chi^2$ of the stars of the Musca field analyzed in this work (in section \ref{results}). As it can be seen, a  $\chi^2 = 6.0$ separates reasonably good those stars belonging to the Musca cloud and those showing unexpected high polarization. Therefore, a $\chi^2 \le 6.0$  would include a reasonable sample with real polarization. This test was applied in other fields with same result. However, a different cutoff for  $\chi^2$ can be selected by the user of {\sc solvepol} either interactively or a priori before running the code.

{\sc Solvepol} doesn't correct for Ricean bias affecting values with $P/\sigma_P \leq 3.0$ \citep{Clarke:1986}.  The users of {\sc solvepol} must be aware of this and correct their results. The results presented in this work were obtained with $P/\sigma_P\ge 5.0$ and $10.0$, when indicated; therefore,  they are not affected by Ricean bias - only values with $P/\sigma_P \leq 3.0$ are affected. 

\subsection{Astrometry calibration}\label{astrometry:calibration}

{\sc Solvepol} uses the {\sc astrometry.net} software \citep{Lang:2010}  to calibrate the astrometry of the image being analyzed to get the accurate celestial coordinates of the stars. The pipeline masks out one of the components  of the doublet (always the most right component) of each star emerging from the calcite Savart prism on the first waveplate position's image, to get a `normal' image of the  sky.  Figure \ref{FieldDoublet} shows the original image from the first waveplate position (left), and after applying the mask (right). For this case the right-up component of each star have been masked. The masked regions are replaced with the mode sky value of the image. The size of the mask is equal to the aperture that gives the lowest uncertainty in $P$. This `normal' image is  processed by the {\sc astrometry.net} software\footnote[3]{Alternatively, in the case that the astronomy.net software is not installed in the running computer, the  `normal'  image or a list produced by {\sc solvepol} of positions of the detected stars sorted by brightness  can be used directly in the web version of {\sc astrometry.net}.}.

The {\sc astrometry.net} software finds, in the masked  image, the relative positions of nearby groupings of 3 or 4 stars on the field, and finds their corresponding triples or quadruplets in the 2MASS and Tycho-2 catalogs. Each grouping generates a candidate matching sky, and {\sc astrometry.net} finds the true matching view. After an alignment of the image and the catalog, {\sc astrometry.net} calculates  the astrometric solution that the pipeline includes in the masked image header. This image/header information is then used to translate from pixels to RA and Dec of the detected sources. 

The mean accuracy in the position that we calculated by comparing the position of standard stars (see section \ref{stdstarsSec}) with that of the literature is $\pm 0.44$ arcsec. This accuracy allows to search the stars in the Guide Star Catalog version 2.3  \citep[GSC~v2.3.][]{Lasker:2008} to perform the magnitude calibration.


\subsection{Photometry and magnitude calibration}

New photometric measurements are applied to both the ordinary and extraordinary images within an aperture with diameter 2 times the mean FWHM and the default annulus in {\sc solvepol} to estimate the sky background of 10 pixel radius and 10 pixels width. The mean FWHM is estimated of at most 30 of the brightest stars in the field, in the first waveplate position's image. If the FWHM is greater than 10 pixels, an aperture of 10 pixels in radius is used for the photometry.  We define the measured flux,  $F_m$,  as the sum of the ordinary and extraordinary fluxes. For our observations taken in the V-band, the pipeline obtains the visual magnitude:

\begin{equation}
V=m_z - 2.5  log (F_m) 
\end{equation}

\noindent  where $m_z$ is the zero point magnitude calibrated with the GSC~v2.3 \citep{Lasker:2008}.  This catalog goes down to 18th magnitude\footnote[4]{The pipeline code can be modified to calibrate with any catalog available in the VizieR database over the web.}. 
 We define $m_z$ as the mean of the absolute difference (to avoid that the sum is zero) between our measured magnitude, $m_{{\rm m,i}}$, and the optical magnitude reported in the GSC~v2.3, $m_{{\rm c},i}$, for at most the 25 brightest stars in the field, i.e.,  $m_z\! =\! \frac{1}{k}\sum^{k}_{i=1}\! \mid \!m_{{\rm m},i}\! -\! m_{{\rm c},i}\!\mid$,  where $k$ is at most 25. If there are no stars with measured magnitude in the catalog GSC~v2.3, $m_z$ is set to 0 (no calibration).

\section{Results}\label{results}

We conducted pilot studies to test our new pipeline {\sc solvepol}. These include the analysis of ten standard stars (essential for calibration), the field centered on HD110984 (a well studied field in our group), and the already published 1997 observational campaign on Musca \citep[][hereafter \citetalias{Pereyra:2004}]{Pereyra:2004}.   We  fixed $F/\sigma_s \geq 5.0$ for this present work -- a limit for a confident source detection.

\subsection{Standard stars}\label{stdstarsSec}

The analysis of polarized standard stars is essential to check the veracity of our results. For the 10 standard stars shown in Table \ref{stdstars}, we preset polarization and magnitude results using {\sc solvepol} and compare these with reported values from the literature. The standard stars were observed during different observational campaigns, most of them in the 1997 Musca's campaign \citepalias{Pereyra:2004}. 

\begin{table*}[]
\begin{center}
\caption{Standard stars analyzed. Our measurements and those from the literature (last line for each star) are shown. RA, Dec and $V$ are from the GSC v2.3 catalog \citep{Lasker:2008}; $P$ is from: 
 (a)  \citet{Bastien:1988},
 (b) \citet{Serkowski:1975} (the error is not provided in this reference) and 
 (c) \citet{Turnshek:1990}.
`---' indicates that the position could not be obtained by the pipeline because of the lack of reference stars on the filed (see Section \ref{stdstarsSec} for details), and then the magnitude could not be calibrated by {\sc solvepol}.}  
\label{stdstars}
\begin{tabular}{*{7}{l}}
\tableline
\tableline
\multicolumn{1}{c}{Star} &   \multicolumn{1}{c}{RA J2000}       &  \multicolumn{1}{c}{Dec J2000}        &  \multicolumn{1}{c}{$V$} &   \multicolumn{1}{c}{$P$} &  \multicolumn{1}{c}{Obs.} & Ref.  \\ 
	                              &   &  &  \multicolumn{1}{c}{(mag)}&   \multicolumn{1}{c}{(\%)}  &   \multicolumn{1}{c}{date}\\ 
\tableline
HD80558 &$09  \   18  \    42.44$  &  $ -51   \   33    \  37.69$ & $5.29 \pm   0.20$&$ 3.25\pm 0.06$ & 1997-04-09&\multicolumn{1}{c}{...} \\
		&$09   \   18   \   42.36    $&$ -51  \   33  \    38.34 $&$5.94\pm0.01$ & $3.11\pm 0.01$ &\multicolumn{1}{c}{...}  & \multicolumn{1}{c}{a}  \\
HD84810 &  \multicolumn{1}{c}{---} &\multicolumn{1}{c}{---}&\multicolumn{1}{c}{---} &$1.67\pm  0.07$& 1997-04-09 & \multicolumn{1}{c}{...} \\ 
	        &$09   \   45     \ 14.81$&   $  -62  \    30   \   28.44$ & $3.69\pm0.03$ & $1.58\pm 0.01$ &\multicolumn{1}{c}{...} & \multicolumn{1}{c}{a}\\
HD100623 &\multicolumn{1}{c}{---} & \multicolumn{1}{c}{---}& \multicolumn{1}{c}{---} & $0.04\pm  0.06$  & 1997-04-10 &\multicolumn{1}{c}{...} \\ 
		&$11  \   34    \  29.49$   &$  -32  \    49  \    52.81$& $6.05\pm0.01$ & $ 0.02$ &\multicolumn{1}{c}{...} & \multicolumn{1}{c}{b} \\
HD110984 &$12  \    46   \   44.84 $   &$ -61  \    11     \ 11.91$  & $9.19 \pm 0.18$  & $5.64 \pm0.05$ & 2001-03-02&\multicolumn{1}{c}{...}    \\ 
		&$12   \   46    \  44.84$&     $-61  \    11   \   11.47 $  & $8.49\pm    0.16$ & $ 5.79\pm 0.04$  &1997-04-12&\multicolumn{1}{c}{...} \\ 
  		& $12  \    46   \   44.82$&     $-61  \    11 \     11.47$ & $8.55  \pm  0.16$ & $5.86\pm 0.03$ & 1997-04-13&\multicolumn{1}{c}{...} \\ 
                & $12 \     46    \  44.84  $  & $-61     \ 11   \   11.58  $   & $9.04\pm0.02$  &$ 5.70\pm 0.01$ &\multicolumn{1}{c}{...} & \multicolumn{1}{c}{c} \\ 
HD111613	 &$12   \   51   \   18.05$&     $-60   \   19  \    47.79 $ & $5.47 \pm   0.17$ &  $3.20\pm 0.04$  &1997-04-09&\multicolumn{1}{c}{...} \\ 
		& $12   \   51   \   17.98$&     $-60    \  19  \    47.24 $& $5.77\pm0.01$ & $3.06 \pm 0.01$ &\multicolumn{1}{c}{...}  &\multicolumn{1}{c}{a}  \\
HD126593 &$ 14   \   28   \   50.87$&    $ -60  \    32    \  24.94$ & $  8.60 \pm   0.29$&$ 4.94\pm 0.06$ &  2001-03-02&\multicolumn{1}{c}{...} \\ 
                 & $14     \ 28   \   50.87$&     $-60   \   32    \  25.12$ &  $8.70\pm0.01$ & $5.02\pm 0.01$  &\multicolumn{1}{c}{...}  &\multicolumn{1}{c}{c} \\ 
HD147084 & \multicolumn{1}{c}{---} & \multicolumn{1}{c}{---} & \multicolumn{1}{c}{---} & $ 3.79 \pm 0.35 $&	1997-04-10&\multicolumn{1}{c}{...}  \\ 
		 &$16     \ 20   \   38.18$&     $-24   \   10    \  09.55 $&$4.66\pm0.01$ & $4.17\pm 0.01$&\multicolumn{1}{c}{...}  &\multicolumn{1}{c}{c} \\
HD154445 & \multicolumn{1}{c}{---} & \multicolumn{1}{c}{---} & \multicolumn{1}{c}{---} & $3.79   \pm 0.05$ & 1997-04-10&\multicolumn{1}{c}{...}  \\ 
		 & $17    \   05   \   32.26$&       $-00   \  53    \  31.44$ & $5.65\pm0.01$ & $3.80\pm 0.08$ &\multicolumn{1}{c}{...}  & \multicolumn{1}{c}{c}  \\
HD155197 & $ 17   \   10  \    15.74$&     $ -04 \     50  \    03.62$  & $9.60\pm    0.08$ & $4.54\pm 0.09$ & 1997-04-12&\multicolumn{1}{c}{...}   \\ 
 		& $17    \  10   \   15.75$&     $ -04   \   50   \   03.67$& $ 9.57\pm0.03$ & $ 4.63\pm  0.02$ &\multicolumn{1}{c}{...} & \multicolumn{1}{c}{c} \\
HD298383 &$ 09    \  22   \   29.77$&     $-52    \  28  \    57.65 $& $ 10.78 \pm   0.72$&$ 5.17\pm 0.06 $  & 2001-03-01&\multicolumn{1}{c}{...}  \\ 
		&$ 09    \  22   \   29.77$&     $-52    \  28  \    57.36$& $10.74  \pm  0.37$ & $5.43\pm 0.04$ & 1997-04-13&\multicolumn{1}{c}{...}  \\  
		&$09     \ 22   \   29.76$&     $-52    \  28  \    57.25 $& $10.78 \pm   0.34 $ & $ 5.38\pm 0.06$ & 1997-04-12&\multicolumn{1}{c}{...} \\ 
                & $ 09     \ 22 \     29.77$&     $-52   \   28 \      57.25 $&  $9.75\pm0.03$ & $5.23\pm 0.01$ &\multicolumn{1}{c}{...}  & \multicolumn{1}{c}{c} \\ 
\tableline
\end{tabular}
\end{center}
\end{table*}

Although  the standard stars are used to get the zero-angle calibration of the observations, the disposition of the telescope relative to the sky or to different positions of the waveplate inside the polarimeter is not the same between observing  campaigns.  Hereby, the angle $\theta$ is meaningless to compare between stars observed in different campaigns, and it is not tabulated in Table \ref{stdstars}. The position of HD100623, HD147084 and HD154445 could not be solved because less than three stars fall in their field -- at least three stars are needed to solve the astrometry -- therefore, its RA and Dec are not tabulated for its comparison with the literature.

\begin{figure*}
\centering{
\includegraphics[angle=90,width=12cm]{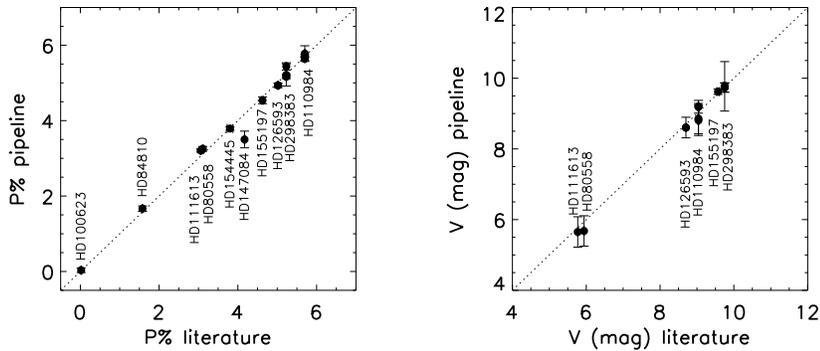}
\caption{Left: Polarization of the standard stars reduced with {\sc solvepol} versus the polarization in the literature (see Table \ref{stdstars}). Right: $V$ magnitudes reduced by {\sc solvepol} versus the $V$ magnitude from the GSC~v2.3 \citep{Lasker:2008}. Dashed line represents the equality. 
 The plotted error bars are $1\sigma$.}
\label{PvsPlit}
}
\end{figure*}

Figure \ref{PvsPlit}, shows plots comparing $P$ and $V$ of the standard stars calculated with {\sc solvepol} and  from the literature. These results are consistent to well within $2\sigma$. The $P$ that deviate by more corresponds to HD147084: our  $P$ is lower than the reported in \citet{Turnshek:1990} by a factor of $0.9$.
Because we do not see in Figure \ref{PvsPlit} a systematic displacement from the line representing the equality, the scatter can be attributed to climatological differences along the observation (clouds, cirrus, etc.).


\subsection{HD110984 and its field}\label{HD110984:field}

\begin{figure*}
\centering{
\includegraphics[width=7cm]{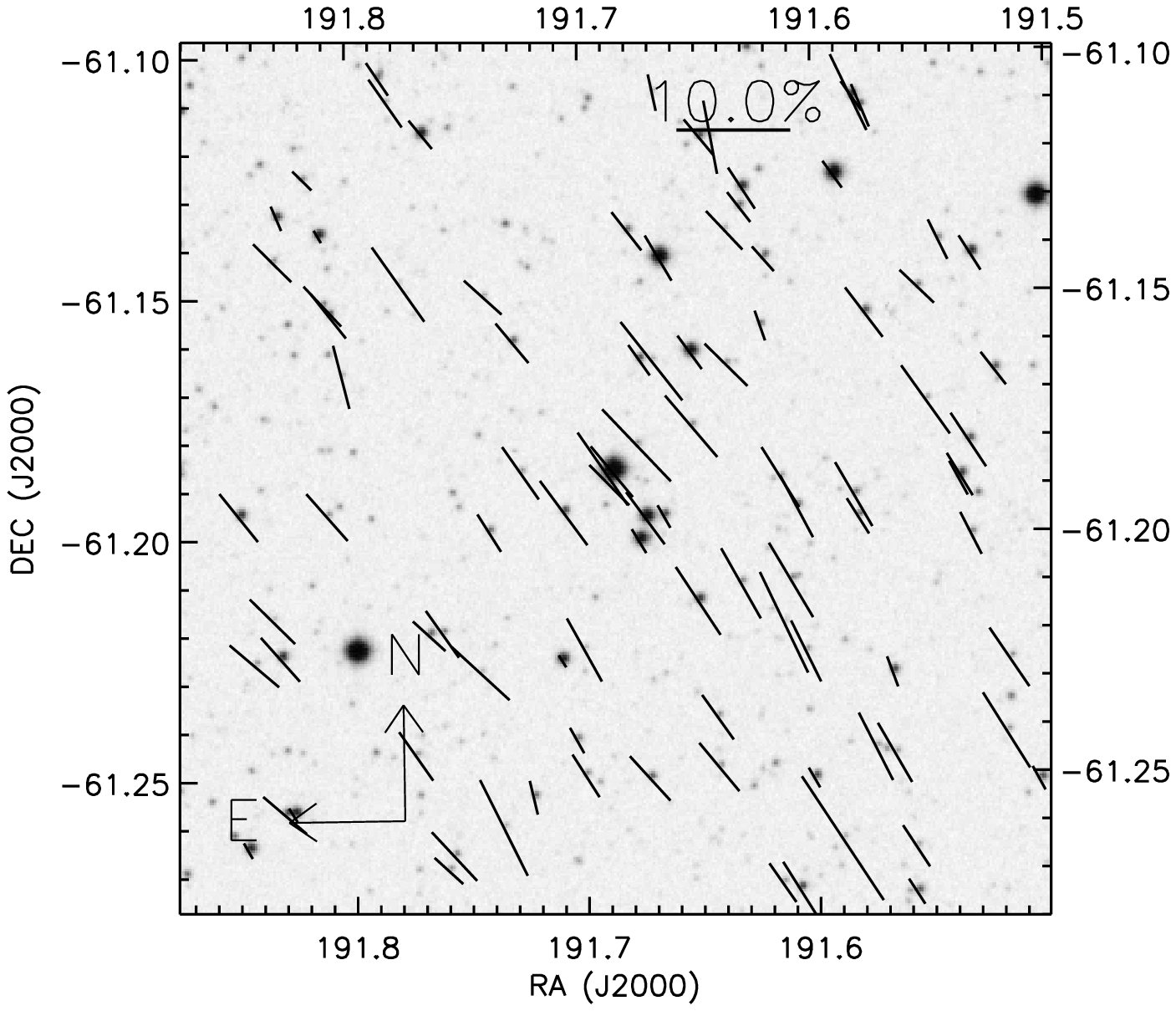} 
\includegraphics[width=7cm]{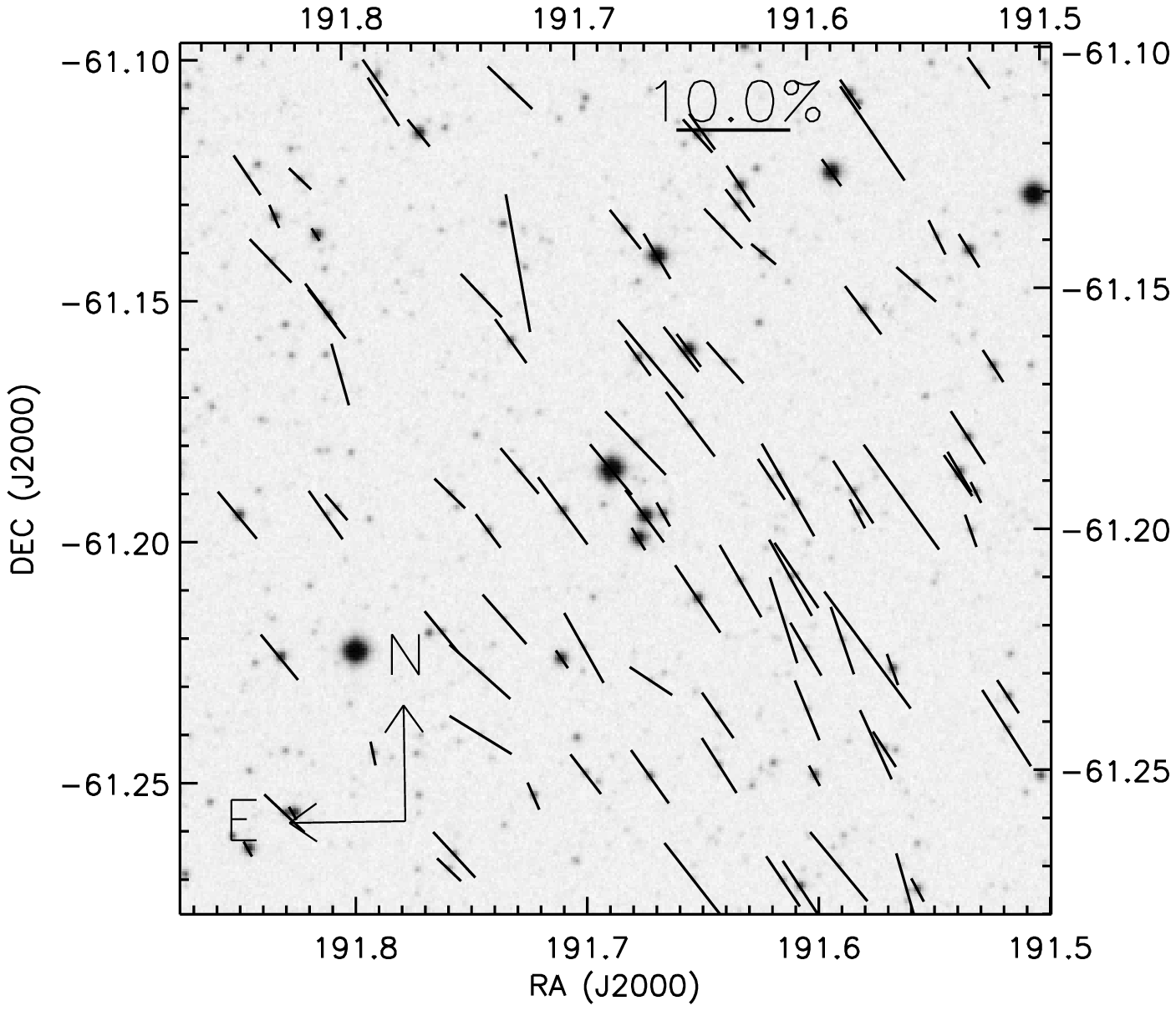} 
\caption{Left:  Polarization map of the field centered on the standard star HD110984 reduced with the {\sc pccdpack} routines. Right:  Same field reduced with {\sc solvepol}. $F/\sigma_s\geq 5.0$ and $P/\sigma_P\ge 5.0$. 
}
\label{pmaps}
}
\end{figure*}

We have analyzed the field around HD110984. Figure~\ref{pmaps} shows the polarization maps of the field centered on HD110984 solved by {\sc solvepol} and by  {\sc pccdpack}. The polarization (length of the lines) and position angle (measured from North to East) are indicated. Both maps will lead to the same conclusion about the distribution of the polarization on the field of HD110984. However, there are some discrepancies. Around RA $191.85$ and Dec $-61.22$, {\sc pccdpack} solves three stars, while  {\sc solvepol} solves one star only, which is the brightest star of the group of three stars.  Moreover, in the center of the image, corresponding to the standard star HD110984,  {\sc solvepol} solves precisely for HD110984, while {\sc pccdpack} solves three sources where clearly  there is no stars but the spikes of HD110984. This examples indicate  that {\sc solvepol} is producing  reliable results.

\begin{figure*}
\centering{
\includegraphics[angle=90,height=4cm]{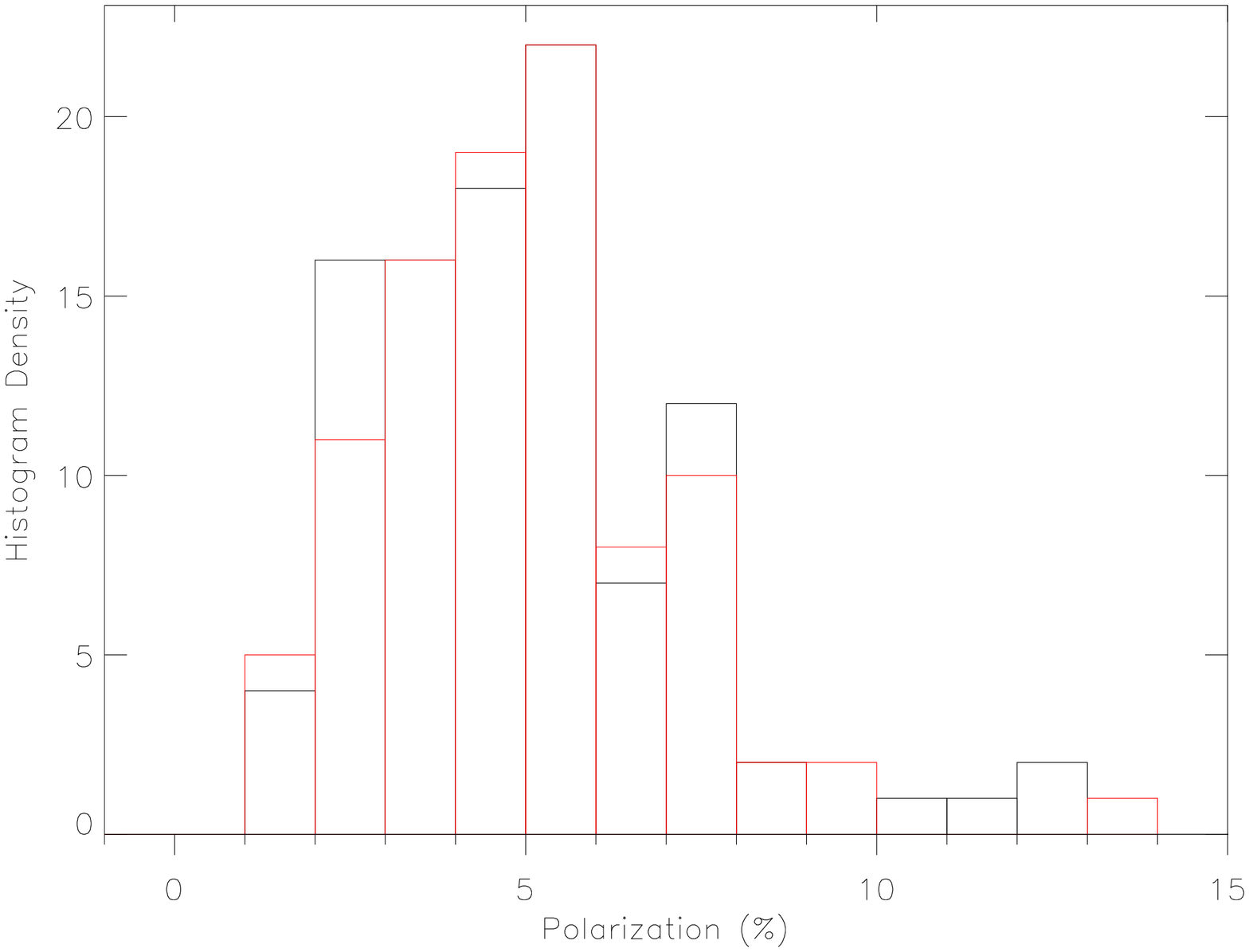}
\includegraphics[angle=90,height=4cm]{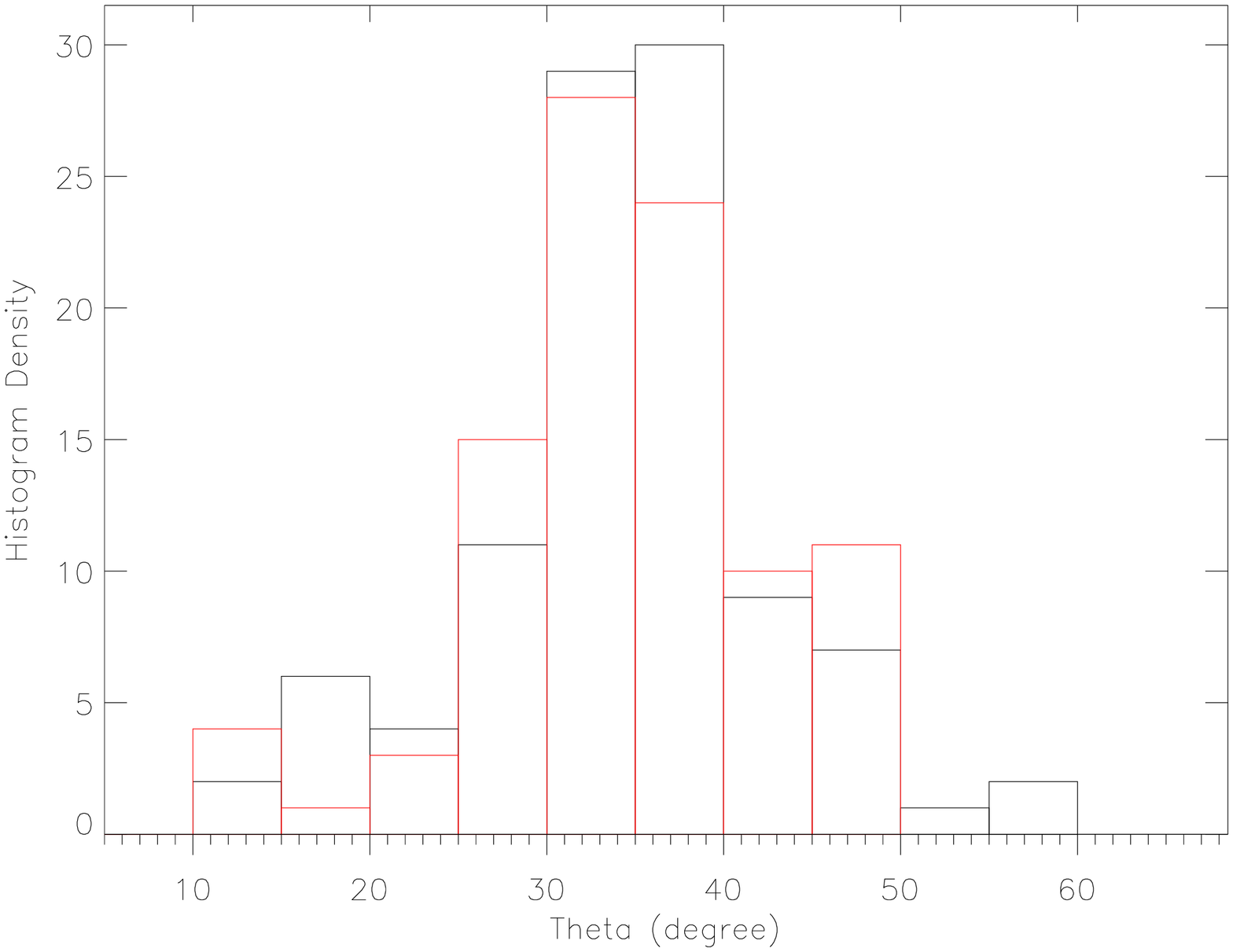}
\includegraphics[angle=90,height=4cm]{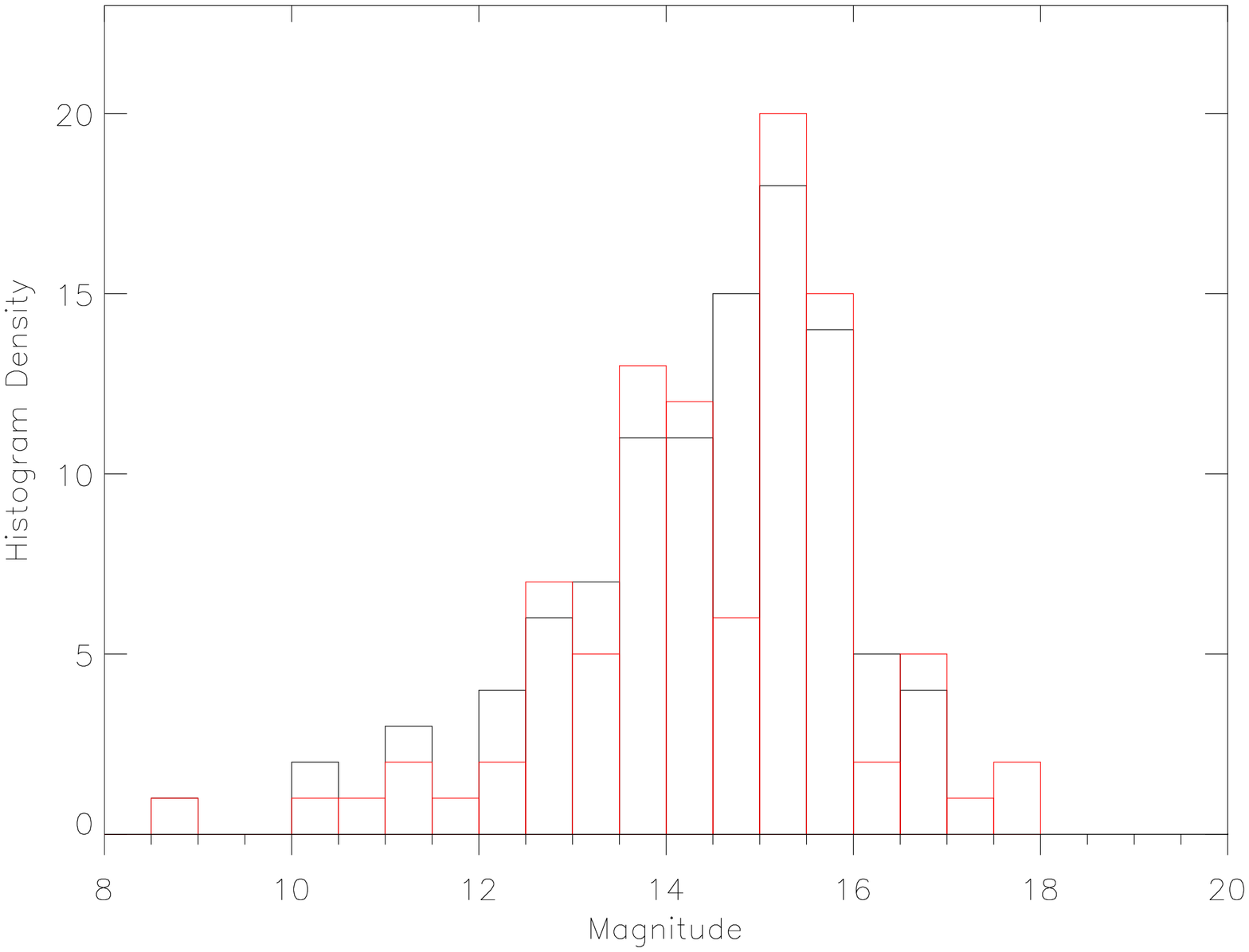}
\caption{Distributions of $P$, $\theta$ and $V$  measured by {\sc solvepol} (black) and the {\sc pccdpack} routines (red) of the field centered on HD110984. These distributions correspond to $F/\sigma_s\geq 5.0$ and $P/\sigma_P\ge 5.0$.} 
\label{hd110distributions}
}
\end{figure*}

Figure \ref{hd110distributions} shows the distributions of $P$, $\theta$ and $V$ for the HD110984  field measured with {\sc solvepol} and the {\sc pccdpack} routines for comparison.  We fixed $F/\sigma_s \geq 5.0$ and $P/\sigma_P\ge 5.0$ for this analysis. The distributions of $P$, $\theta$ and $V$  produced by both routines are quite similar. In particular, the distributions of  $P$ and $\theta$ indicate that both routines lead to the same conclusion about the polarimetric properties on the field of HD110984.  Therefore, both routines give polarimetric properties that will lead to the same conclusions when analyzing the same field.

\begin{figure*}
\hspace{-1cm}\includegraphics[height=18cm,angle=90]{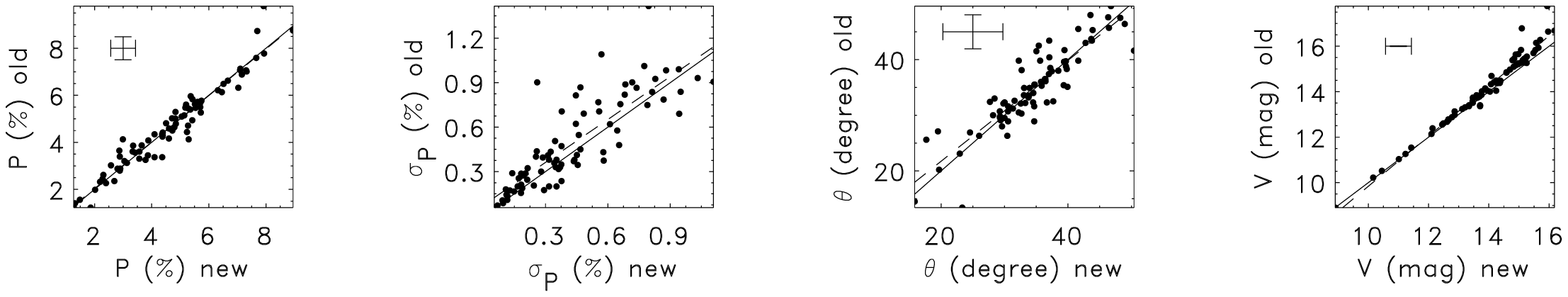} 
\caption{From left to right: $P$, $\sigma_P$, $\theta$ and $V$ measured with {\sc solvepol} (new) and those measured by the previous {\sc pccdpack} routines (old) on HD110984's field. Solid  line represents the equality; Dashed line is a linear fit for comparison. The mean $1\sigma$ error bar is indicated.}
\label{compare}
\end{figure*}

In Figure \ref{compare} we compare $P$, uncertainty in the polarization ($\sigma_P$), $\theta$ and  $V$ using {\sc solvepol} versus those using the  {\sc pccdpack} routines. For the same stars found by {\sc solvepol} and the {\sc pccdpack} routines,  it can be seen that both routines produce the same results with each other to within $3\sigma$.  Clearly the $V$-error bar given by {\sc pccdpack} is underestimated because it is unexpectedly lower than the mean error  reported in the catalog GSC~v2.3  for this field, which is $\pm 0.43$.   {\sc Solvepol}, on the other hand, takes the mean error of the magnitude reported in the GSC~v2.3 catalog as a lower limit of its $V$-error, ie., if  {\sc solvepol} measures an  $V$-error  lower than the mean $V$-error of the magnitudes reported in the GSC~v2.3,  {\sc solvepol} takes the mean $V$-error of the magnitude  of the catalog as the error. About the mean error in $\theta$ in Figure \ref{compare}, the mean error in theta is $\pm4.73$~degrees and $\pm3.06$~degrees for  {\sc solvepol} and {\sc pccdpack}, respectively. A difference as low as $1.67$ degrees indicates consistency in $\theta$.

\begin{figure*}
\centering{
\includegraphics[width=7cm]{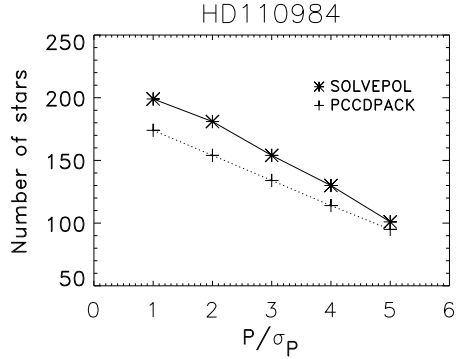}
\caption{Number of detected stars of the field of HD110984 versus $P/\sigma_P$ measured by {\sc solvepol} (asterisk signs) and by {\sc pccdpack} (plus signs).  $F/\sigma_s$ is fixed to $\geq 5.0$. Dotted and solid lines are simply joining the points. Notice that {\sc solvepol}  gives catalogs with more stars than  {\sc pccdpack}, and the number of sources converge towards greater values of  $P/\sigma_P$. For instance for  $P/\sigma_P\ge 5.0$ the difference in number of detected stars is only 6.}
\label{compareN}
}
\end{figure*}

In Figure \ref{compareN} we compare the number of sources detected by {\sc solvepol} and {\sc pccdpack}. The number of sources detected by {\sc solvepol} tends to be higher than the number of sources detected by the {\sc pccdpack} routines for low values of $P/\sigma_P$ (with $F/\sigma_s \geq 5.0$).  However, the number of sources detected by both routines  converge towards greater values of $P/\sigma_P$.  For instance, for  $P/\sigma_P=3.0$, the number of stars detected by {\sc solvepol} is 154, and the number of sources detect by the {\sc pccdpack} routines is 134 -- a difference of 20 detected stars --, while for $P/\sigma_P=5.0$ (usual ratio for a confident measurement) the difference is just 6 stars (101 and 95  detection by {\sc solvepol} and the {\sc pccdpack} routines, respectively). Therefore, our new pipeline seems to be producing more complete catalogs.

\subsection{Musca Dark Cloud data} 

A comparison of our results with those reported in \citetalias{Pereyra:2004} of the well studied field of the Musca Dark Cloud is presented in this section. We analyze all the fields observed in the 1997 campaign. The observational details are described in \citetalias{Pereyra:2004}.  The source detection limits  used in {\sc solvepol} were set to $F/\sigma_F \geq 5.0$ and $P/\sigma_P \geq 5.0$, same limits adopted by \citetalias{Pereyra:2004}, but using the {\sc pccdpack} routines.

\begin{figure*}
\hspace{-1cm}\includegraphics[height=18cm,angle=90]{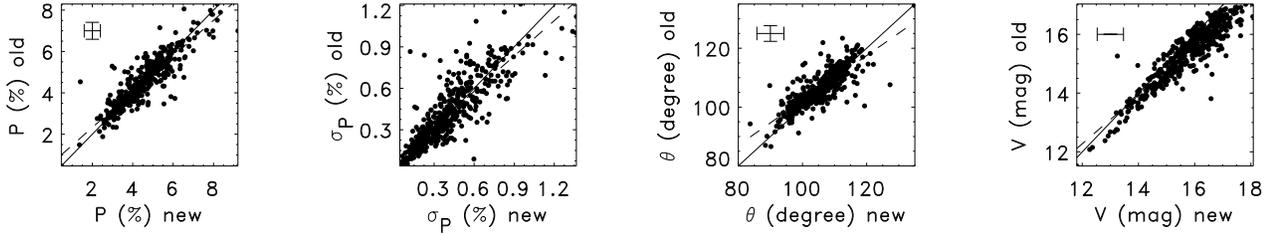}
\caption{From left to right: $P$, $\sigma_P$, $\theta$ and $V$ measured with {\sc solvepol} (new) and those reported in VizieR obtained by \citetalias{Pereyra:2004} using the {\sc pccdpack} routines (old). $F/\sigma_s\geq 5.0$ and $P/\sigma_P\ge 5.0$. Solid line represents the equality; dashed line is a linear fit for comparison.  The mean $1\sigma$ error bar is indicated. There is no magnitude error bars reported in \citetalias{Pereyra:2004}. 
}
\label{compareMusca}
\end{figure*}

In Figure \ref{compareMusca} the  $P$, $\sigma_P$, $\theta$ and  $V$ calculated by {\sc solvepol}, and those from the catalog of \citetalias{Pereyra:2004} are presented. As it can be seen, $P$, $\sigma_P$, $\theta$ and $V$ are all consistent with each other, with some degree of scattering within $3\sigma$. The catalogs produced by {\sc solvepol} were processed with {\sc merge}, therefore, there are no repeated stars from the overlapping field regions and the weighted mean of $P$, $\theta$ and $V$ of those overlapped stars were calculated (see section \ref{description:pipeline}). On the other hand, the catalog (reported in VizieR) of \citetalias{Pereyra:2004} includes two entries for stars in overlapping regions, increasing the total number of results. The uncertainty in $V$ is not reported by \citetalias{Pereyra:2004}, and thus we cannot compare our uncertainties against theirs.

While the error bars in $P$ calculated by {\sc solvepol} and {\sc pccdpack}  indicated in  Figure \ref{compareMusca} are consistent, the error bar of $\theta$ calculated by {\sc solvepol}  is slightly bigger than the error bar in $\theta$ calculated by {\sc pccdpack}.  The mean error in $\theta$ is $\pm4.23$ and $\pm2.66$ for {\sc solvepol} and {\sc pccdpack}, respectively. This gives a  difference of only 1.57 degrees, which indicates that  {\sc solvepol} and {\sc pccdpack} are in close agreement.

\begin{figure*}
\centering{
\includegraphics[angle=90,height=4cm]{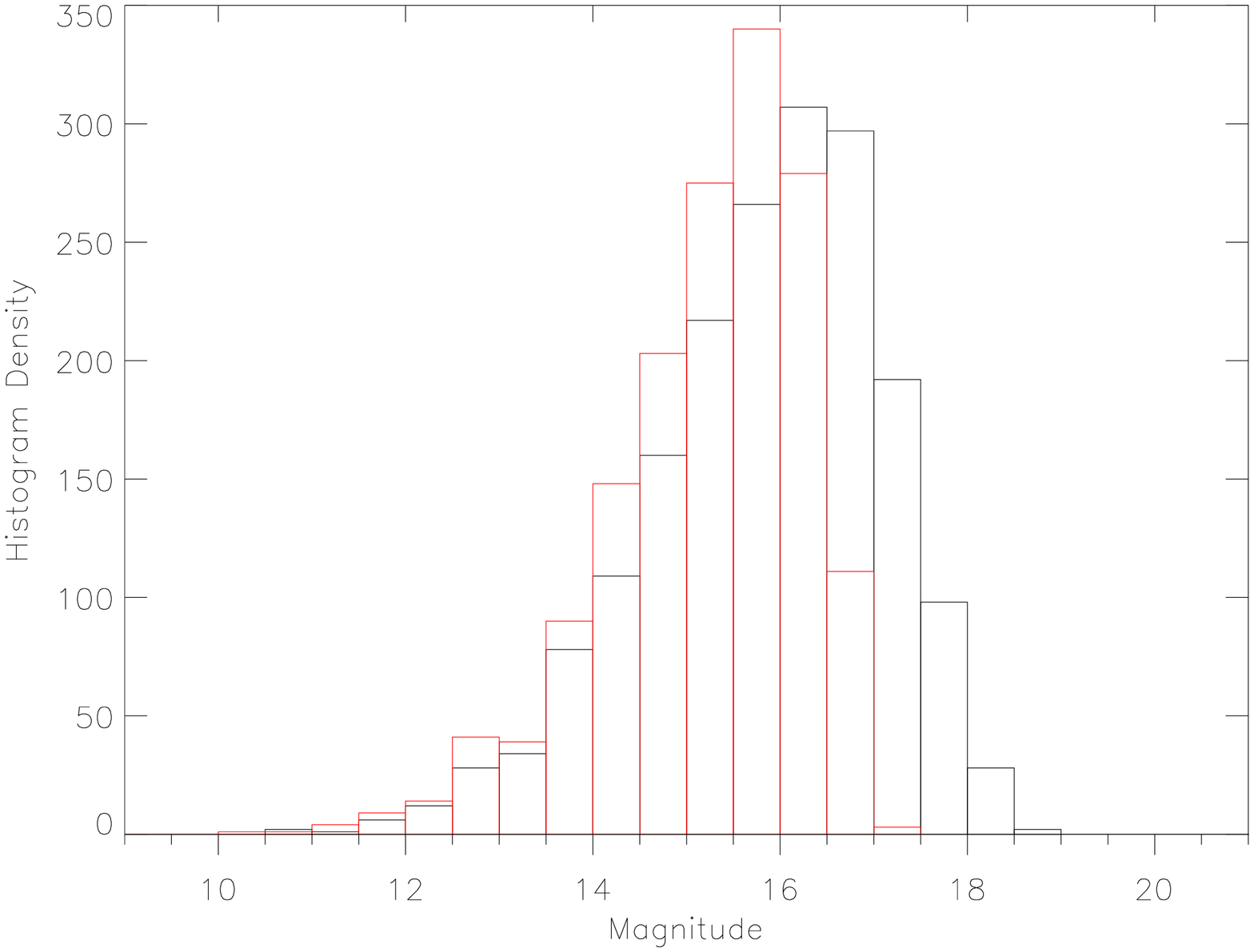}
\includegraphics[angle=90,height=4cm]{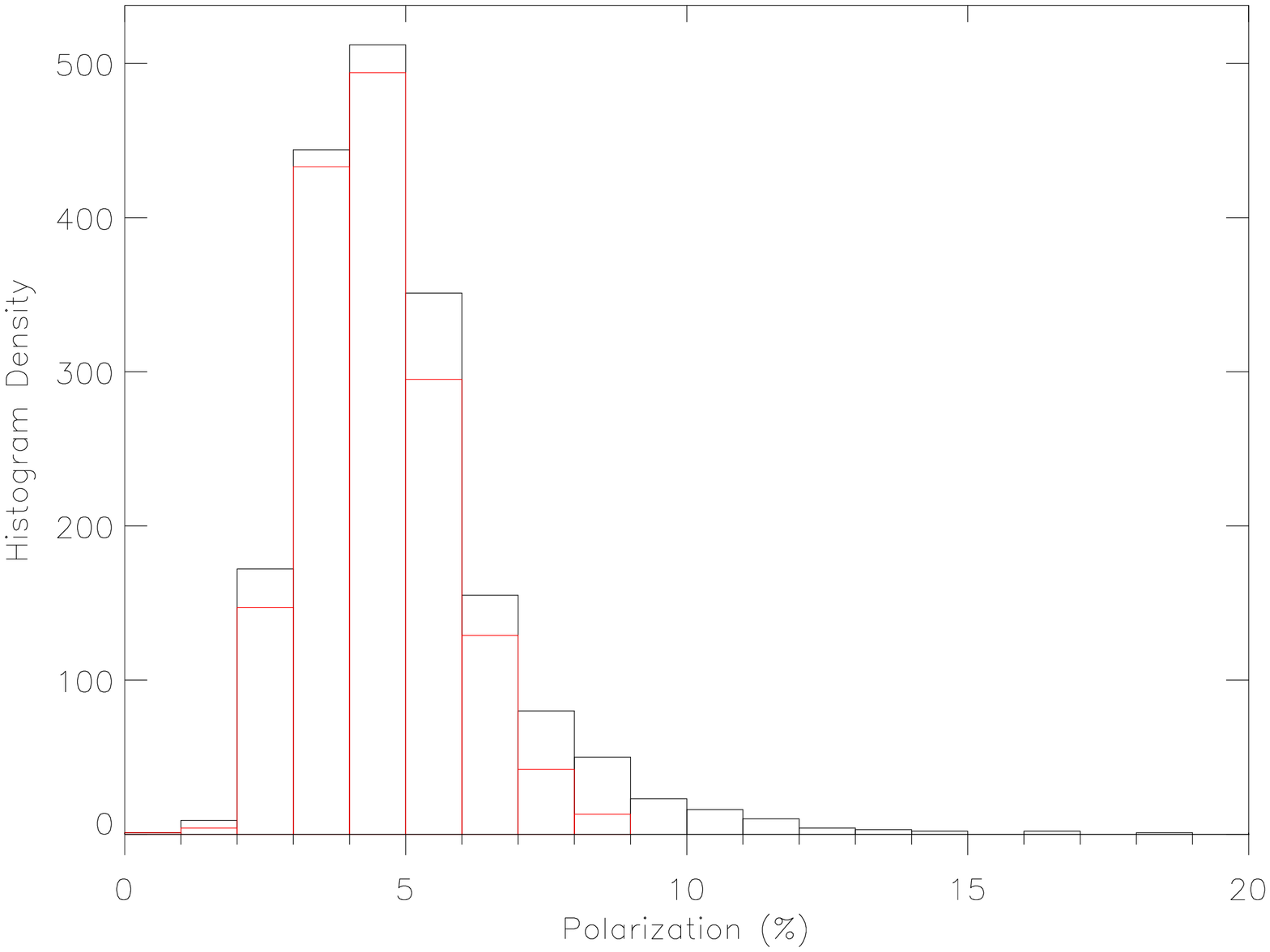}
\includegraphics[angle=90,height=4cm]{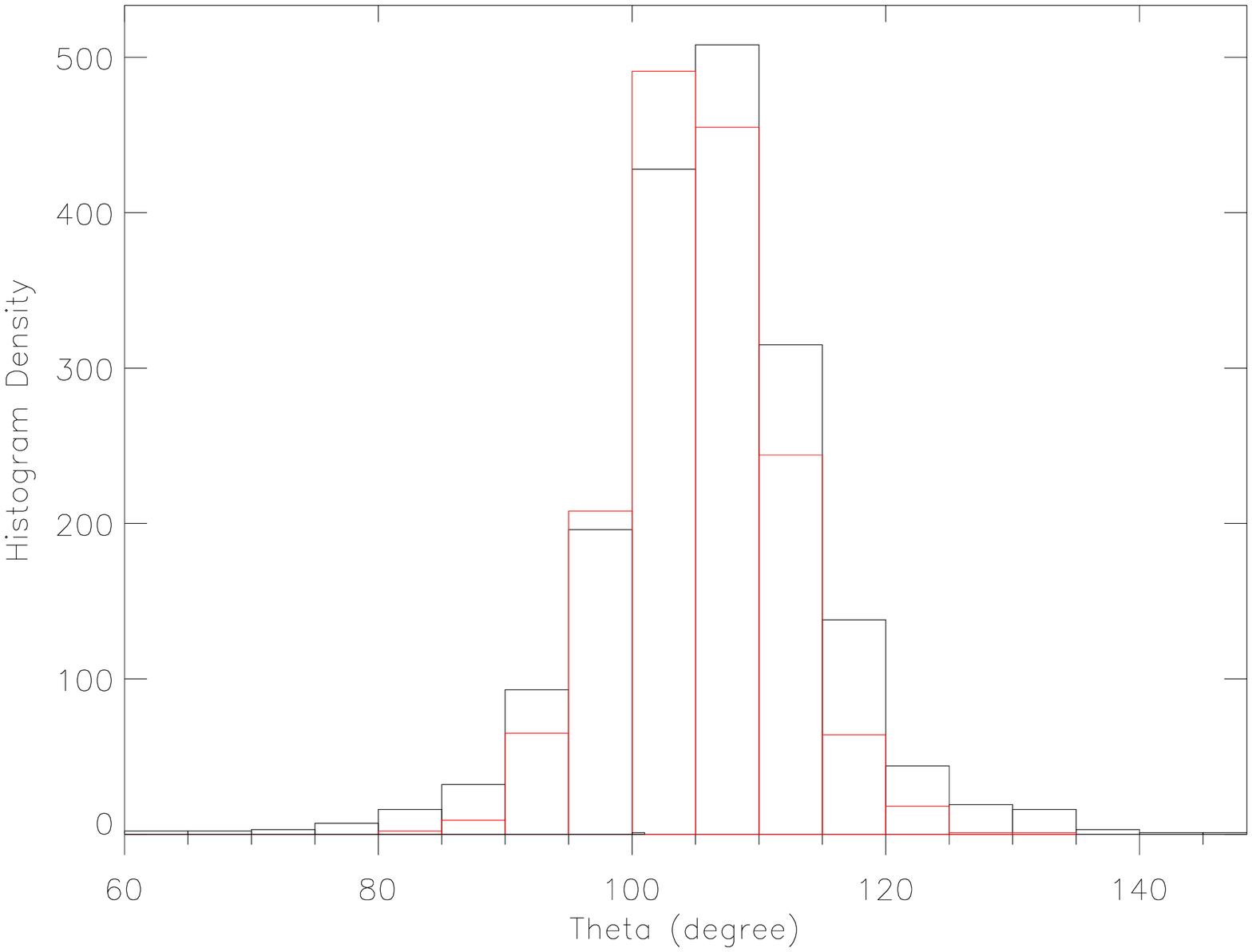}
\caption{Distributions of  $P$,  $\theta$ and $V$ of Musca measured by {\sc solvepol} (black) and by the {\sc pccdpack} (red). The selection is $F/\sigma \geq 5.0$  and $P/\sigma_P \geq 5.0$.
 \label{muscaDistCase1}}
}
\end{figure*}

\begin{figure*}
\centering{
\includegraphics[angle=0,height=5cm]{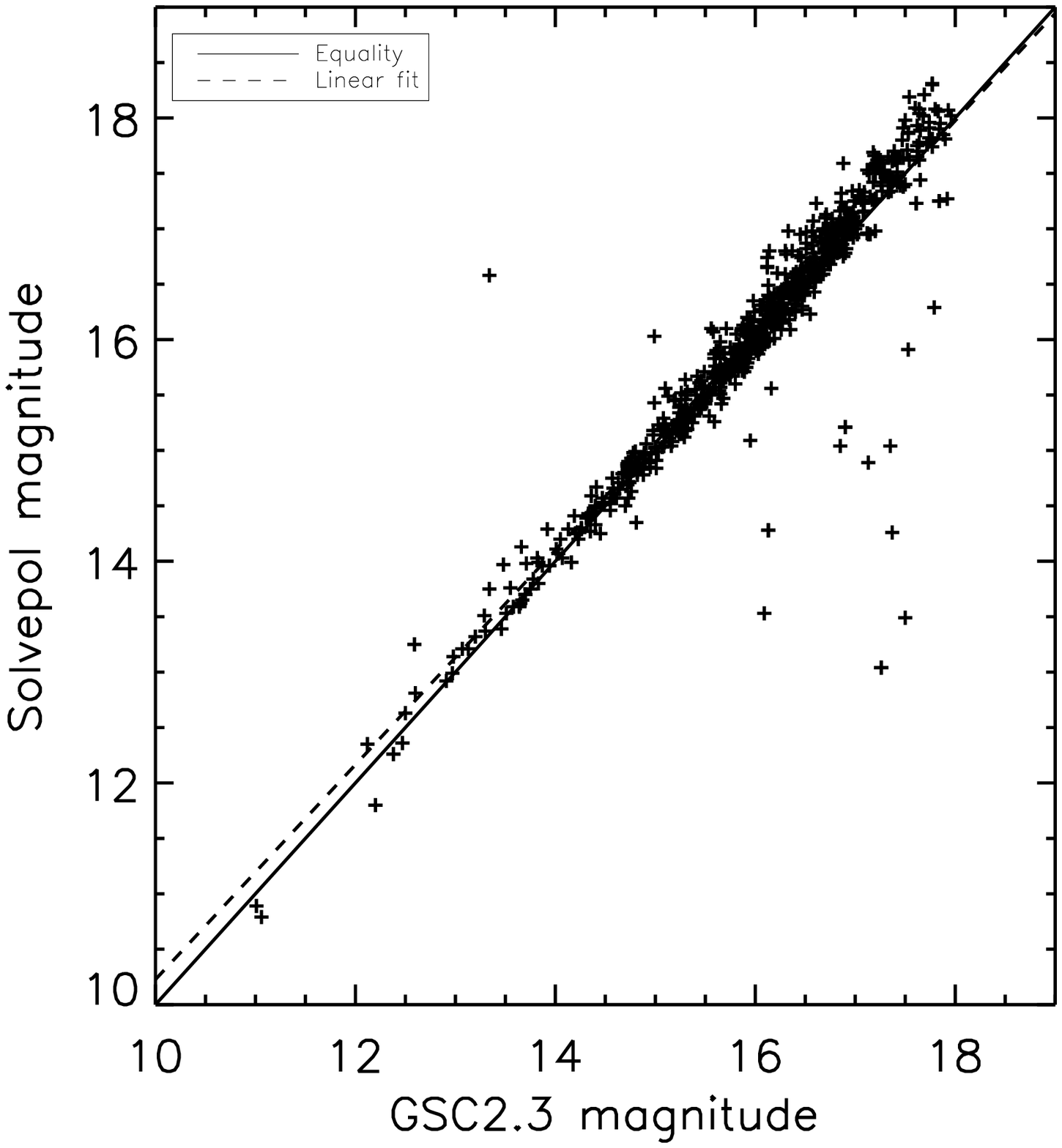}%
\includegraphics[angle=0,height=5cm]{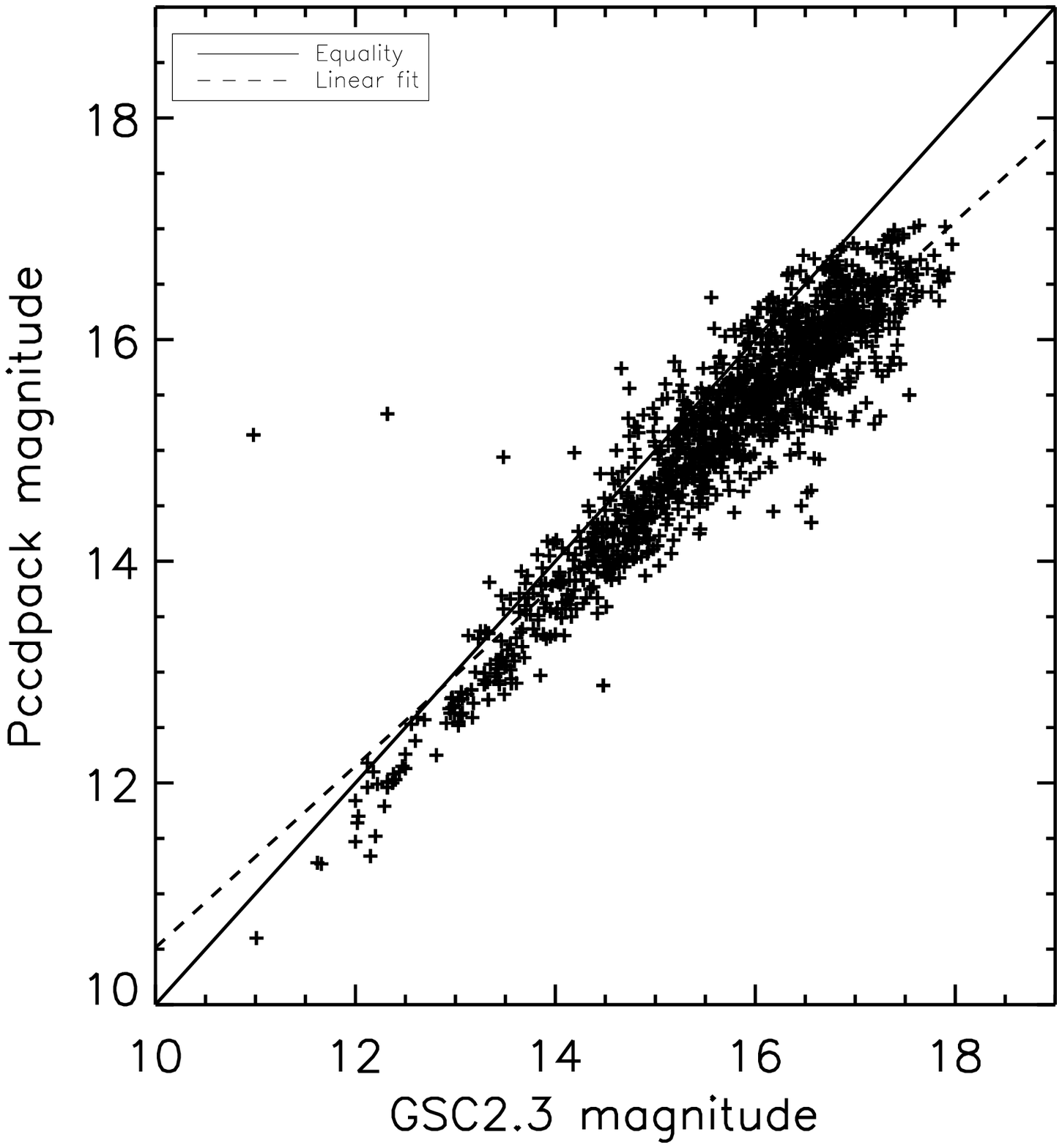}%
\caption{
Magnitudes measured by {\sc solvepol} (left) and by the {\sc pccdpack} (right) vs the magnitude reported in the GSC~v2.3. Solid line represents the equality; dashed line is a linear fit for comparison.  
 \label{compareGSC}}
}
\end{figure*}

The distributions of  $P$, $\theta$, and $V$  for Musca are shown in Figure \ref{muscaDistCase1}.  The sample reduced with {\sc solvepol} seems to be complete in magnitude up to $17$ mag. This is $1.5$ magnitudes deeper than the $15.5$ mag reported in \citetalias{Pereyra:2004} using {\sc pccdpack}.  However, a systematic shift of the estimated magnitudes is clearly seen. To know which magnitudes are offset, we compare both magnitudes with the magnitudes of the GSC~v2.3, show in Figure \ref{compareGSC}. As it can be seen, the magnitudes estimated by  {\sc solvepol} are closer to the equality line with the magnitude of the GSC~v2.3 than those estimated by  {\sc pccdpack}: the  {\sc solvepol} magnitudes fall very close to the equality line, while the  magnitudes from {\sc pccdpack} is offset bellow the equality line. Therefore, the magnitude calibration performed by {\sc solvepol} is more accurate with the GSC~v2.3.

Table \ref{table:fields} shows the field numbers \citepalias[we have adopted the same notation used in][]{Pereyra:2004}, its central  position calculated by our pipeline, the number of detected stars, and the weighted mean (as derived in \citetalias{Pereyra:2004}) of $P$, $\theta$  and  $V$ with their respective uncertainties calculated by {\sc solvepol}, and those from \citetalias{Pereyra:2004} for comparison.  We adopted the same limit of $P/\sigma_P \geq 10.0$  adopted by \citetalias[][their table 5]{Pereyra:2004}, for a direct comparison. 

\begin{table*}
\begin{center}
\caption{
Fields of Musca reduced by {\sc solvepol} using the values $P/\sigma_P \geq 10.0$ and $F/\sigma \geq 5.0$. 
 RA and Dec are the central position of the field.
 N is he number of detected sources in each field. 
 Weighted mean of $P$, $\theta$  and  $V$ and its standard deviation following the same procedure detailed in \citetalias{Pereyra:2004}.
 Below, same fields of Musca reported by \citetalias{Pereyra:2004} and using the same $P/\sigma_P$ and $F/\sigma$ limits. } 
\label{table:fields}
\begin{tabular}{c*{7}{l}}
\tableline
\tableline
Field & \multicolumn{1}{c}{RA J2000}           & \multicolumn{1}{c}{Dec J2000} &  \multicolumn{1}{c}{N} & \multicolumn{1}{c}{mean $P$} & \multicolumn{1}{c}{mean $\theta$} & \multicolumn{1}{c}{mean $V$}  \\
         & \multicolumn{2}{c}{Central Region}  &                                                   & \multicolumn{1}{c}{(\%)} & \multicolumn{1}{c}{($^\circ$)} & \multicolumn{1}{c}{(mag)}  \\
\tableline
 03 & $12\      35\       05.73$&    $-71 \     00\       30.80 $& 60 & $3.16 \pm 0.02$ & $101.455 \pm 1.84$& $15.46 \pm 1.22$ \\
 05 & $12\      30\       06.91$&    $-71\      00\       35.20 $& 48 & $3.82 \pm 0.02$ & $105.507 \pm 1.58$ & $15.50 \pm 1.06$\\
 08 & $12\      27\       36.99$&    $-71\      08\       40.13 $& 46 & $3.45 \pm 0.03$ & $107.602 \pm 2.09$ & $14.94 \pm 0.92$\\
 11 & $12\      30\       10.15$&    $-71\      24\       38.34 $& 35 & $3.83 \pm 0.03$ & $104.835 \pm 1.63$ & $15.41 \pm 1.24$\\
 13 & $12\      25\       13.27$&    $-71\      24\       40.64 $& 64 & $5.09 \pm 0.02$ & $111.129 \pm 1.62$ & $15.47 \pm 1.08$\\
 14 & $12\      27\       43.58$&    $-71\      32\       35.64 $& 33 & $2.56 \pm 0.03$ & $105.917 \pm 1.74$ & $15.63 \pm 1.15$\\
 16 & $12\      27\       50.15$&    $-71\      41\       00.17 $& 61 & $4.68 \pm 0.03$ & $100.026 \pm 1.63$ & $15.73 \pm 1.00$\\
 20 & $12\      25\       12.13$&     $-72\     00\       56.48 $& 67 & $4.39 \pm 0.02$ & $107.791 \pm 1.71$ & $15.29 \pm 1.11$\\
 22 & $12\      25\       18.97$&     $-72\     09\       19.19 $& 63 & $4.05 \pm 0.02$ & $109.961 \pm 1.73$ & $14.92 \pm 1.16$\\
 24 & $12\      19\       58.77$&     $-72\     09\       33.49 $& 87 & $4.73 \pm 0.02$ & $105.628 \pm 1.63$ & $14.91 \pm 1.18$\\
 26 & $12\      20\       06.32$&     $-72\     17\       26.06 $& 75 & $5.08 \pm 0.02$ & $104.501 \pm 1.60$ & $14.86 \pm 1.08$\\
 27 & $12\      22\       33.51$&     $-72\     25\       16.68 $& 40 & $4.85 \pm 0.03$ & $116.587 \pm 1.62$ & $14.87 \pm 0.98$\\
 29 & $12\      19\       57.42$&     $-72\     33\       32.76 $& 53 & $4.44 \pm 0.02$ & $102.888 \pm 1.64$ & $15.10 \pm 1.39$\\
 30 & $12\      18\       02.47$&     $-72\     33\       39.58 $& 71 & $3.24 \pm 0.02$ & $109.563 \pm 1.67$ & $14.41 \pm 1.12$\\
 34 & $12\      37\       56.59$&     $-70\     41\       10.27 $& 54 & $2.74 \pm 0.02$ & $106.526 \pm 1.69$ & $14.42 \pm 1.20$\\

\tableline
 03&$12 \ 35 \ 27.35$ &$-71 \ 00 \ 31.29$&55 &3.15 $\pm$0.02& 102.31 $\pm$1.79& $      15.45\pm     1.00$\\
 05&$12 \ 30 \ 28.02$ &$-71 \ 00 \ 34.25$&54 &3.80 $\pm$0.02& 106.17 $\pm$1.62& $      15.47\pm     1.01$\\
 08&$12 \ 27 \ 58.53$ &$-71 \ 08 \ 35.56$&48 &3.52 $\pm$0.03& 107.38 $\pm$1.97& $      15.50\pm     1.06$\\
 11&$12 \ 30 \ 28.59$ &$-71 \ 24 \ 34.25$&34 &3.88 $\pm$0.03& 105.95 $\pm$1.55& $      15.34\pm     1.11$\\
 13&$12 \ 25 \ 29.15$ &$-71 \ 24 \ 36.75$&65 &4.93 $\pm$0.02& 111.53 $\pm$1.53& $      15.43\pm     0.97$\\
 14&$12 \ 27 \ 59.05$ &$-71 \ 32 \ 35.56$&34 &3.15 $\pm$0.03& 106.96 $\pm$1.67& $      15.37\pm     1.10$\\
 16&$12 \ 27 \ 59.23$ &$-71 \ 40 \ 35.56$&62 &4.67 $\pm$0.03& 102.71 $\pm$1.70& $      15.47\pm     0.97$\\
 20&$12 \ 05 \ 29.89$ &$-72 \ 00 \ 36.75$&64 &4.56 $\pm$0.03& 105.13 $\pm$1.71& $      15.42\pm     0.98$\\
 22&$12 \ 25 \ 30.06$ &$-72 \ 08 \ 36.75$&65 &3.92 $\pm$0.02& 107.74 $\pm$1.76& $      15.43\pm     0.97$\\
 24&$12 \ 20 \ 15.82$ &$-72 \ 08 \ 38.87$&78 &4.70 $\pm$0.02& 102.45 $\pm$1.57& $      15.48\pm     1.00$\\
 26&$12 \ 20 \ 15.95$ &$-72 \ 16 \ 38.87$&68 &5.03 $\pm$0.02& 101.81 $\pm$1.60& $      15.44\pm     1.00$\\
 27&$12 \ 22 \ 53.25$ &$-72 \ 24 \ 37.87$&49 &5.17 $\pm$0.03& 113.36 $\pm$1.65& $      15.48\pm     1.05$\\
 29&$12 \ 20 \ 16.23$ &$-72 \ 32 \ 38.87$&52 &4.51 $\pm$0.02& 99.72   $\pm$1.70& $      15.50\pm     1.03$\\
 30&$12 \ 17 \ 39.05$ &$-72 \ 32 \ 39.74$&69 &3.25 $\pm$0.02& 107.20 $\pm$1.64& $      15.45\pm     1.00$\\
 34&$12 \ 38 \ 03.74$ &$-70 \ 40 \ 29.55$&49 &2.67 $\pm$0.02& 104.24 $\pm$1.74& $      15.48\pm     1.05$\\

\tableline
\end{tabular}
\end{center}
\end{table*}

\begin{figure*}
\centering{
\includegraphics[width=8.cm]{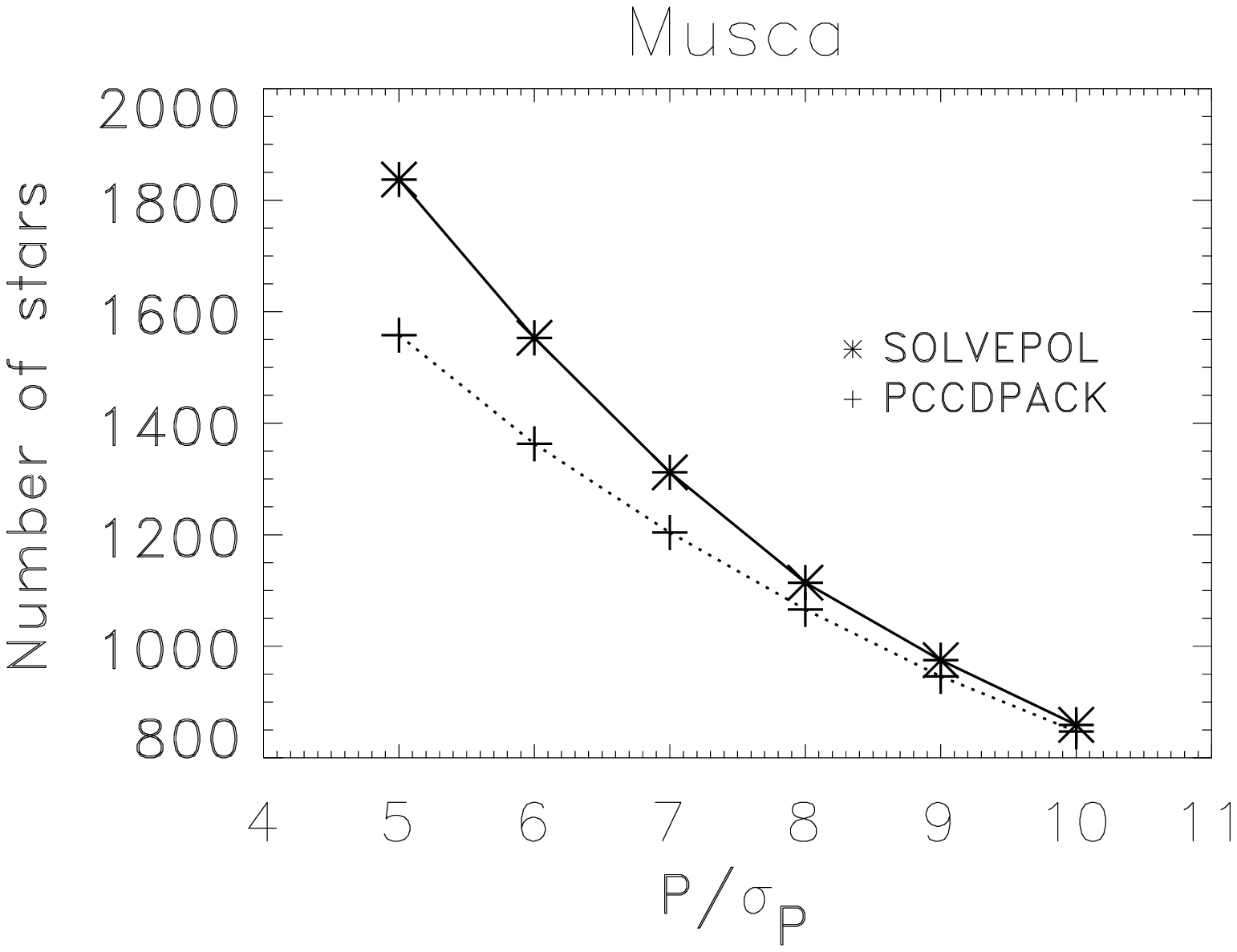} 
\includegraphics[width=8.cm]{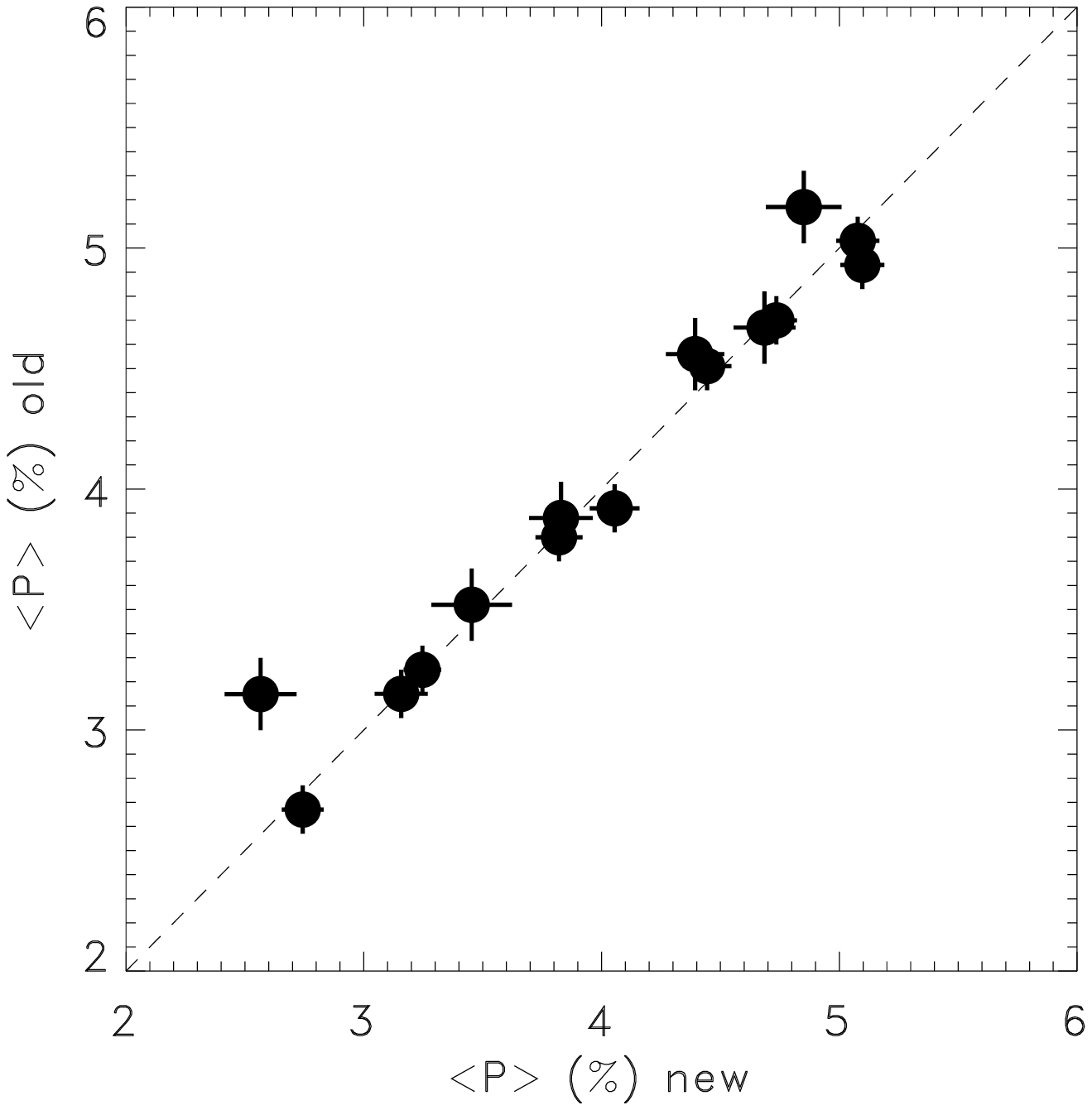}\\
\includegraphics[width=8.cm]{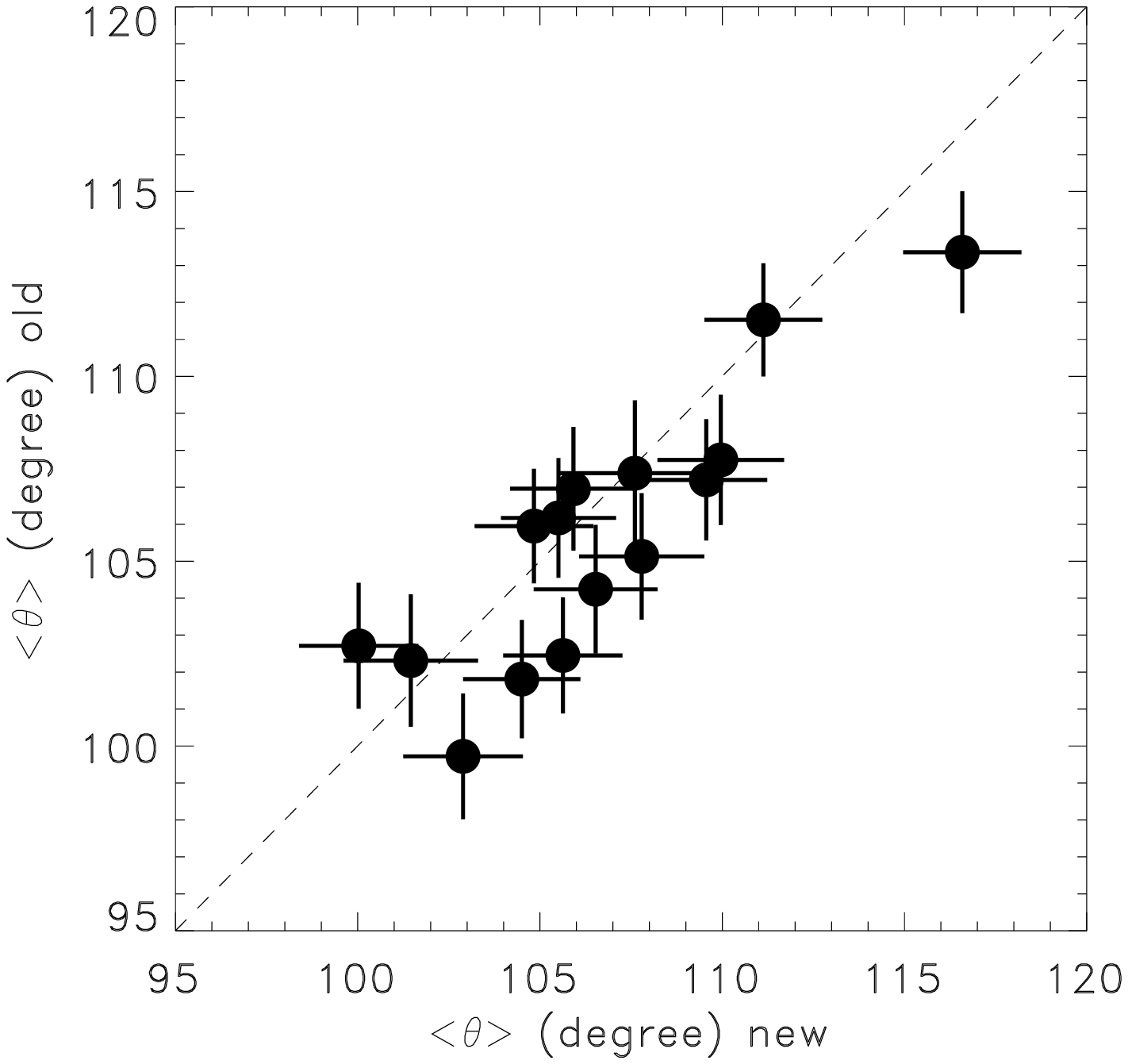} 
\includegraphics[width=8.cm]{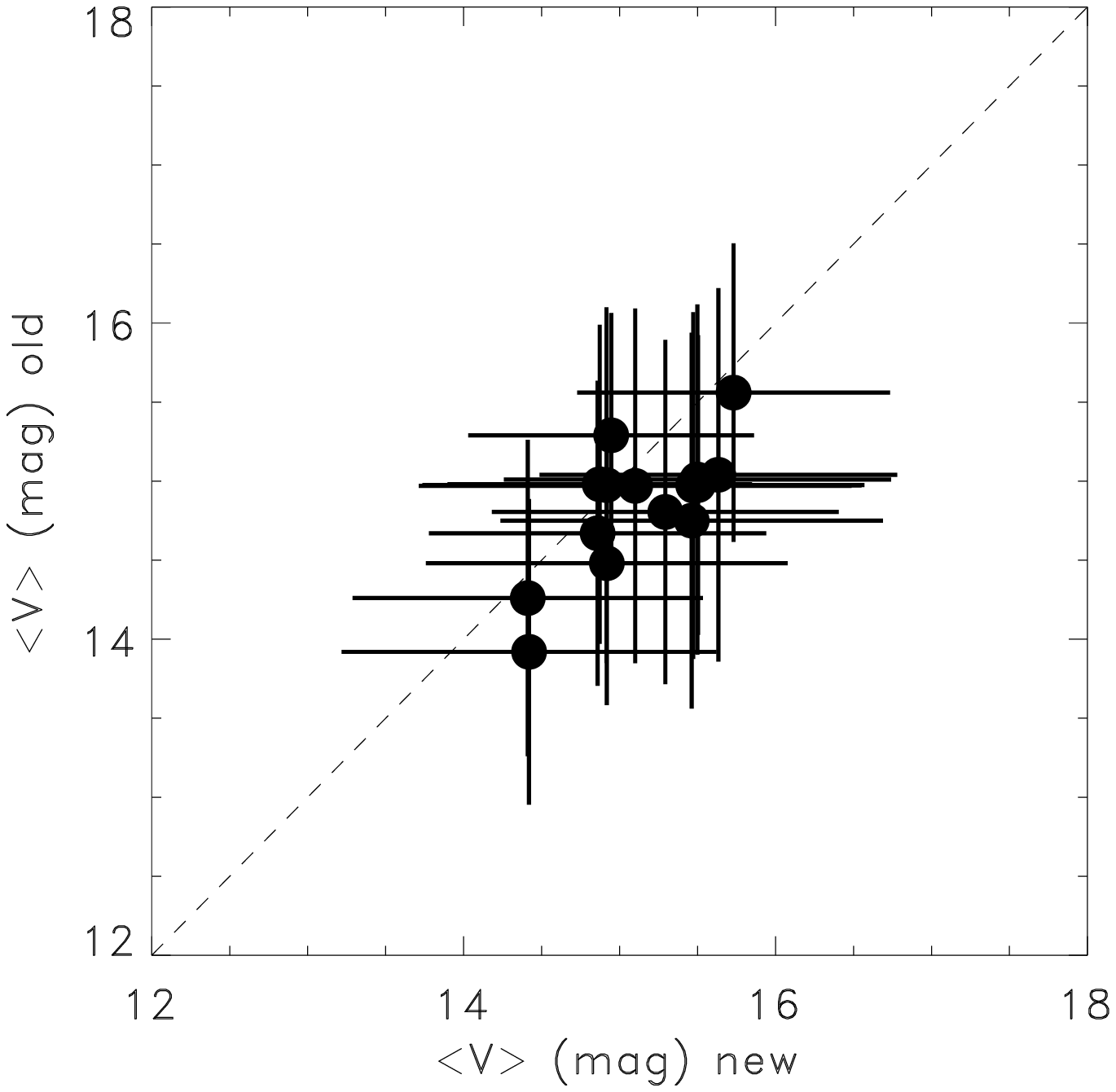} 
\caption{
Top Left: number of detected stars of Musca versus $P/\sigma_P$ measured with {\sc solvepol}  (asterisk signs) and the number of sources reported in VizieR obtained by \citetalias{Pereyra:2004} using the {\sc pccdpack} routines (plus signs). $F/\sigma_s$ is fixed to $\geq 5.0$. Clearly the number of sources converge towards greater values of  $P/\sigma_P$.  For the next plots $F/\sigma_s\geq 5.0$ and $P/\sigma_P \geq 10.0$.
Top Right:  weighted mean values of $P$  measured with {\sc solvepol} (new) and weighted mean values of $P$ reported in VizieR obtained by \citetalias{Pereyra:2004} (old). The errors have been amplified by 5 for an easier inspection. 
Bottom Left: weighted mean values of $\theta$ measured with {\sc solvepol} (new) and weighted mean values of $\theta$ reported in VizieR obtained by \citetalias{Pereyra:2004} (old).
Bottom Right: mean values of $V$ measured with {\sc solvepol} (new) and mean values of $V$ reported in VizieR obtained by \citetalias{Pereyra:2004} (old).
The error bars are our weighted standard deviation from the mean and those reported by \citetalias{Pereyra:2004}, which we took directly from their table. 
 The data of these plots are also reproduced in Table \ref{table:fields}.
}
\label{musca:comparison}
}
\end{figure*}

The total number of stars with  $P/\sigma_P \geq 10.0$  is 857, slightly higher than the 846 sources reported by \citetalias{Pereyra:2004}.  About individual fields, the greatest difference in the number of detected stars is only 9 ({\sc solvepol}  found more) which correspond to field number 24 (see Table \ref{table:fields}). However, as it can be seen in Figure \ref{musca:comparison} (top left plot), that compares the number of sources vs $P/\sigma_P$ ($F/\sigma_F$ fix to 5.0) detected by {\sc solvepol} and the {\sc pccdpack} routines, the number of sources detected by both routines tend to converge for high values of $P/\sigma_P$.

Overall,   mean values of $P$, $\theta$,  and  $V$, tabulated in Table \ref{table:fields}, and plotted in Figure \ref{musca:comparison} for an easier inspection, are consistent with the mean values of $P$, $\theta$  and  $V$ reported by \citetalias[][their table 5, reproduced here in Table \ref{table:fields}]{Pereyra:2004}.

\begin{figure*}
\centering{
\hspace{-1.cm}\includegraphics[height=6.3cm]{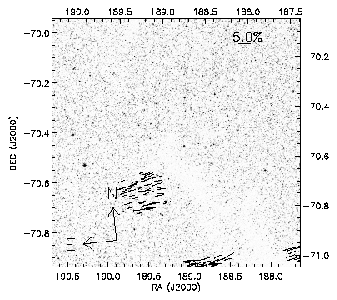}\\
\hspace{-1.cm}\includegraphics[height=6.3cm]{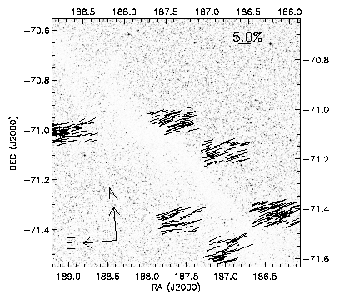}\\
\hspace{-1.cm}\includegraphics[height=6.3cm]{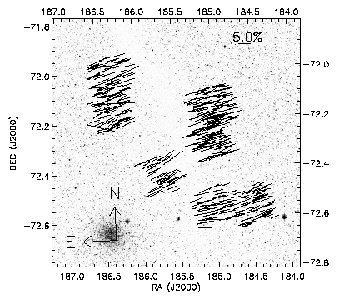}
\caption{Polarization maps of Musca produced with {\sc solvepol} to compare with those in  \citetalias{Pereyra:2004}. The selection is $F/\sigma \geq 5.0$ and $P/\sigma_P \geq 10.0$. \label{pol:maps}}
}
\end{figure*}

Figure \ref{pol:maps} shows the polarization maps on the Musca region of the fields analyzed in this work and tabulated in Table \ref{table:fields}  calculated by {\sc solvepol}. Each vector represents one object. Again, our polarization maps and those reported by  \citetalias{Pereyra:2004} are in agreement.

It is important to highlight that an important difference between {\sc solvepol} and {\sc pccdpack} is that  {\sc solvepol} is completely automatic, a pipeline without human intervention. The results of  {\sc pccdpack}, on the other hand,  have been cleaned  of stars with unexpected polarimetric values by the user.  The results of {\sc solvepol} compared in this work  have not  been post-processed in such a way.  Therefore,  the raw output of {\sc solvepol} is  in reasonable agreement with the post-cleaning output of {\sc pccdpack}.

\section{Comparison between IDL and IRAF photometric routines}\label{photo:section}

An aspect to note is that there are stars in the final catalog by {\sc solvepol}, but are not in the final catalog produced by the {\sc pccdpack} routines, and vice versa. This can be seen in Figure~\ref{pmaps}, which shows the polarization maps of the field of the standard star HD110984 solved by {\sc solvepol} and by  {\sc pccdpack}. 
The difference in the final result could be due to differences in the photometry performed by {\sc solvepol} and {\sc pccdpack};  {\sc solvepol}  is written in {\sc idl}  and  {\sc pccdpack} is written in  {\sc iraf} and {\sc fortran}.

 

\begin{figure*}
\hspace{1cm}
\includegraphics[height=\textwidth,angle=90]{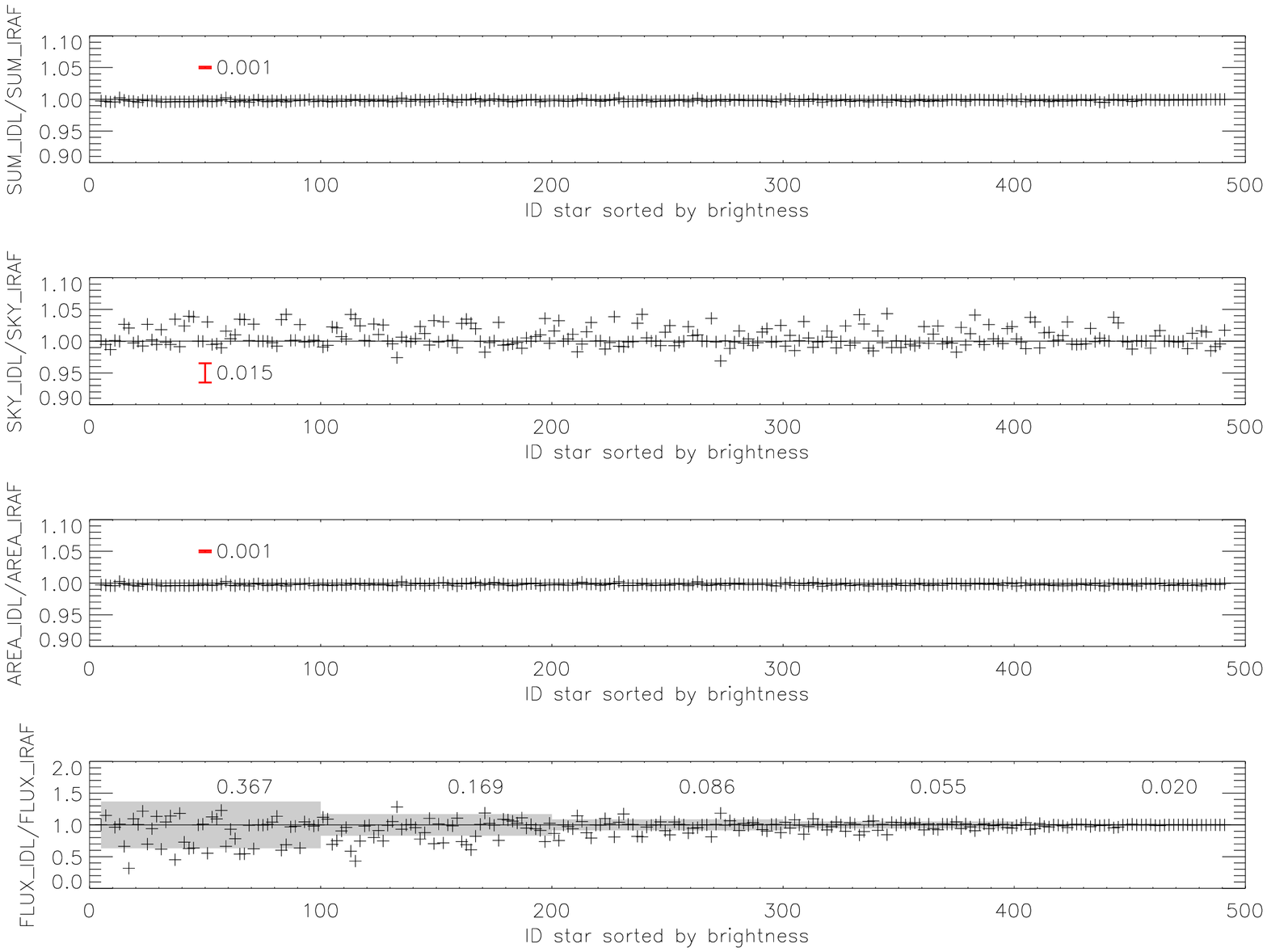}
\caption{Comparison of the photometry performed with {\sc idl} and {\sc iraf} versus an identification number of the stars sorted by brightness (dimmest sources to the left, brightest sources to the right). 
From top to bottom panel:
 (a) the ratio of the total value in counts within the 6 pixels radius aperture  ($SUM$);
 (b) the ratio of the values of the sky within the used annulus  ($SKY$);
 (c) the ratio of the area of the aperture ($AREA$);
 (d) the ratio of the final fluxes ($FLUX$).  
 The standard deviation from the mean and its value is indicated for each plot by the red bar. In the case of the $FLUX$, the standard deviation from the mean were calculated in ranges of ID (indicated by the grey shadowed areas and its value given above each grey area), because the scattering variates with brightness. }
\label{phot:compa}
\end{figure*}

 
To explore this possibility, we performed photometry with {\sc idl} and {\sc iraf} using the same input parameters in the same image. We have used an aperture of 6 pixels radius ($\sim 2\times$FWHM; this aperture size collects $\sim 95\%$ of the light of a star) and the  default annulus  of {\sc solvepol} (inner radius of 10 pixels and 10 pixels width) to perform sky background subtraction, centered on the same position, $X$ and $Y$,  given by {\sc solvepol} (re-centering is turned off). In Figure~\ref{phot:compa} we present the comparison of the photometry performed by the two routines.

The ratio of the $SUM$ shows a scatter from the equality line of around $1\%$.  The standard deviation ($SD$)  from the mean of the $SUM$ ratio is $SD_{SUM}=0.001$. This scatter can be attributable to rounding, meaning that the $SUM$ by {\sc idl} and {\sc iraf} are equal.
As can be seen, there is an order of $1\%$ difference in the $SKY$ values up to $5\%$. The standard deviation for this case is $SD_{SKY}=0.015$. Both {\sc idl} and  {\sc iraf} take the mode of the sky within the annulus (our used annulus has the same size). 
For the $AREA$ of the aperture used to perform the photometry, there is a scatter of $1\%$, attributable to rounding, and the $SD_{AREA}=0.001$.
A difference of order $10\%$ and up to $50\%$ in the $FLUX$  ($= SUM - SKY*AREA$) ratio is clear. The standard deviation from the mean increases to dimmer sources; the $SD_{FLUX}$ goes from $0.020$ to $0.367$  (see the grey shadowed areas in the plot of the $FLUX$ ratio in Figure~\ref{phot:compa}). This indicates that there is greater scatter in flux for fainter sources (left side of the plot) than for bright sources (right  side of the plot).

Although there is $1\%$ scatter in the ratio of the $SUM$s and $AREA$, clearly the main source of scatter in the final $FLUX$ ratio is the difference in the  $SKY$ values measured by {\sc idl} and {\sc iraf}, that  diverge by up to $5\%$. Moreover, the faint stars are more affected by the different value of the SKY than the bright stars.

Hence, the photometry performed by {\sc idl} and {\sc iraf} on the same input image do not always give identical fluxes. The difference in the value of the $SKY$ ($5\%$) gives place to a difference of the $FLUX$ of $\sim 50\%$ for fainter stars, and less that $10\%$ for bright stars. This difference in $FLUX$ in turns is propagated through all the pipeline and leads to the difference observed in the final catalogs. 

\begin{figure*}
\centering{
\includegraphics[height=\textwidth,angle=90]{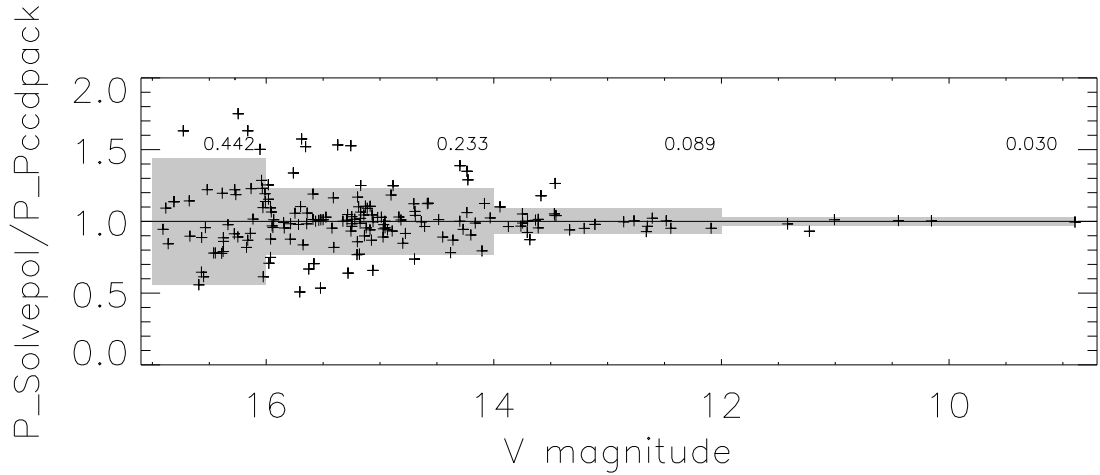}
\caption{Comparison of the polarimetry calculated with {\sc solvepol} and {\sc pccdpack} sorted by magnitude (dimmest sources to the left). The grey shadowed areas are the standard deviation from the mean in ranges of magnitude. The values of the standard deviation for each range are given above each grey area. 
}
\label{pol:compa}
}
\end{figure*}

As expected, the difference in the value of the $FLUX$ affects the final polarization.  In the Figure~\ref{pol:compa} we compare the ratio of the polarization estimated by {\sc solvepol}  and  {\sc pccdpack} versus optical magnitude. The $SD$ from the mean for stars with magnitudes dimmer than $V$~16~mag is $0.442$, greater than the $SD$ for brighter stars (see Figure~\ref{pol:compa}). Therefore,  the value of the sky affects principally fainter stars ($V \geq 13 - 14 $ mag). For bright stars the difference is negligible, and the samples obtained by {\sc solvepol}  and  {\sc pccdpack} are the same. 

We found that the different procedures to estimate the sky within the annulus introduces the main scatter. {\sc idl}  and {\sc iraf} estimates the mode of the sky around the photometric apertures in a slightly different way.  {\sc idl} fits  a Gaussian to the array of the sky values to estimate the mode of the background, eliminating  outlier pixels from the Gaussian distribution to avoid contamination by stars, and the mode of the sky is obtained after 20 iterations. {\sc iraf} estimates the mode of the sky calculating it directly from the sky vector. The different procedures  cause  the flux, and consequently the final polarization  to be not exactly the same, mainly for the fainter stars, and produce the difference in the final catalogs.

\section{Conclusions}\label{conclusion}

We have described {\sc solvepol}, a new pipeline for reducing imaging polarimetry data obtained with the polarimeter IAGPOL, and  that will be used in SOUTH POL.

From the raw images, {\sc solvepol} produces a catalog of the observed field with photometric, polarimetric and astrometric information, as well as plots and tables for an inspection and  analysis of the results. We have compared results from {\sc solvepol} with those from {\sc pccdpack} and found them to be consistent. There is a tendency for {\sc solvepol} to  produce more complete catalogs due to rounding errors and small differences in the codes.  

A clear advantage of {\sc solvepol} over {\sc pccdpack} is  that it produces a final catalog in hours or minutes, while for a user habituated with {\sc pccdpack} it could take hours or  days to produce the final catalog. 

As it stands, {\sc solvepol} is directly applicable to images obtained with birefringent polarizers, such as the simple calcite plate and the calcite Savart prism. It should not be difficult to adapt it to single-image polarizers, such as dichroic polarizers. Another work in progress is adapting the pipeline to NIR imaging polarimetry. The current version of {\sc solvepol} will be made available as a free download at \url{http://www.astro.iag.usp.br/~ramirez/pipeline.html}. It is expected that the optical/IR polarimetry community can contribute to its continuous improvement.





\acknowledgments

EAR  acknowledges FAPESP grant (2012/00185-8) and  the financial support from the Mexican Council of Science and Technology (CONACyT). AMM acknowledges support for his group's activities at IAG from the agencies FAPESP (grant no. 2010/19694-4), CNPq and CAPES. JWD acknowledges funding from the Brazil-U.S. Physics Ph.D. Student \& Post-doc Visitation Program sponsored by the Sociedade Brasileira de F\'isica and the American Physical Society. This research has made use of the VizieR catalog access tool, CDS, Strasbourg, France. The original description of the VizieR service was published in A\&AS 143, 23.

\bibliographystyle{aasjournal}
\bibliography{/Users/eramirez/Dropbox/Articles/bibliography}

\begin{thebibliography}{}
\expandafter\ifx\csname natexlab\endcsname\relax\def\natexlab#1{#1}\fi

\bibitem[{{Adamson} {et~al.}(2005){Adamson}, {Aspin}, {Davis}, \&
  {Fujiyoshi}}]{Adamson:2005}
{Adamson}, A., {Aspin}, C., {Davis}, C., \& {Fujiyoshi}, T., eds. 2005,
  Astronomical Society of the Pacific Conference Series, Vol. 343,
  {Astronomical Polarimetry: Current Status and Future Directions}

\bibitem[{{Bastien} {et~al.}(1988){Bastien}, {Drissen}, {Menard}, {Moffat},
  {Robert}, \& {St-Louis}}]{Bastien:1988}
{Bastien}, P., {Drissen}, L., {Menard}, F., {et~al.} 1988, \aj, 95, 900

\bibitem[{{Bastien} {et~al.}(2011){Bastien}, {Manset}, {Clemens}, \&
  {St-Louis}}]{Bastien:2011}
{Bastien}, P., {Manset}, N., {Clemens}, D.~P., \& {St-Louis}, N., eds. 2011,
  Astronomical Society of the Pacific Conference Series, Vol. 449,
  {Astronomical Polarimetry 2008: Science from Small to Large Telescopes}

\bibitem[{{Clarke} \& {Stewart}(1986)}]{Clarke:1986}
{Clarke}, D., \& {Stewart}, B.~G. 1986, Vistas in Astronomy, 29, 27

\bibitem[{{Clemens} {et~al.}(2012){Clemens}, {Pinnick}, {Pavel}, \&
  {Taylor}}]{Clemens:2012}
{Clemens}, D.~P., {Pinnick}, A.~F., {Pavel}, M.~D., \& {Taylor}, B.~W. 2012,
  \apjs, 200, 19

\bibitem[{{Hecht}(2001)}]{Hecht:2001}
{Hecht}, E. 2001, {Optics 4th edition} (Addison-Wesley Publishing Company)

\bibitem[{{Landsman}(1993)}]{Landsman:1993}
{Landsman}, W.~B. 1993, in Astronomical Society of the Pacific Conference
  Series, Vol.~52, Astronomical Data Analysis Software and Systems II, ed.
  R.~J. {Hanisch}, R.~J.~V. {Brissenden}, \& J.~{Barnes}, 246

\bibitem[{{Lang} {et~al.}(2010){Lang}, {Hogg}, {Mierle}, {Blanton}, \&
  {Roweis}}]{Lang:2010}
{Lang}, D., {Hogg}, D.~W., {Mierle}, K., {Blanton}, M., \& {Roweis}, S. 2010,
  \aj, 139, 1782

\bibitem[{{Lasker} {et~al.}(2008){Lasker}, {Lattanzi}, {McLean}, {Bucciarelli},
  {Drimmel}, {Garcia}, {Greene}, {Guglielmetti}, {Hanley}, {Hawkins},
  {Laidler}, {Loomis}, {Meakes}, {Mignani}, {Morbidelli}, {Morrison},
  {Pannunzio}, {Rosenberg}, {Sarasso}, {Smart}, {Spagna}, {Sturch},
  {Volpicelli}, {White}, {Wolfe}, \& {Zacchei}}]{Lasker:2008}
{Lasker}, B.~M., {Lattanzi}, M.~G., {McLean}, B.~J., {et~al.} 2008, \aj, 136,
  735

\bibitem[{{Magalh\~{a}es} {et~al.}(1984){Magalh\~{a}es}, {Benedetti}, \&
  {Roland}}]{Magalhaes:1984}
{Magalh\~{a}es}, A.~M., {Benedetti}, E., \& {Roland}, E.~H. 1984, \pasp, 96,
  383

\bibitem[{{Magalh\~{a}es} {et~al.}(1996){Magalh\~{a}es}, {Rodrigues},
  {Margoniner}, {Pereyra}, \& {Heathcote}}]{Magalhaes:1996}
{Magalh\~{a}es}, A.~M., {Rodrigues}, C.~V., {Margoniner}, V.~E., {Pereyra}, A.,
  \& {Heathcote}, S. 1996, in Astronomical Society of the Pacific Conference
  Series, Vol.~97, Polarimetry of the Interstellar Medium, ed. W.~G. {Roberge}
  \& D.~C.~B. {Whittet}, 118

\bibitem[{{Magalh{\~a}es} {et~al.}(2005){Magalh{\~a}es}, {Pereyra},
  {Melgarejo}, {de Matos}, {Carciofi}, {Benedito}, {Valentim}, {Vidotto}, {da
  Silva}, {de Souza}, {Faria}, \& {Gabriel}}]{Magalhaes:2005}
{Magalh{\~a}es}, A.~M., {Pereyra}, A., {Melgarejo}, R., {et~al.} 2005, in
  Astronomical Society of the Pacific Conference Series, Vol. 343, Astronomical
  Polarimetry: Current Status and Future Directions, ed. A.~{Adamson},
  C.~{Aspin}, C.~{Davis}, \& T.~{Fujiyoshi}, 305

\bibitem[{{Magalh\~{a}es} {et~al.}(2012){Magalh\~{a}es}, {de Oliveira},
  {Carciofi}, {Costa}, {Dal Pino}, {Diaz}, {Ferrari}, {Fernandez}, {Gomes},
  {Marrara}, {Pereyrac}, {Ribeiro}, {Rodrigues}, {Rubinho}, {Seriacopi}, \&
  {Taylor}}]{Magalhaes:2012}
{Magalh\~{a}es}, A.~M., {de Oliveira}, C.~M., {Carciofi}, A., {et~al.} 2012, in
  American Institute of Physics Conference Series, Vol. 1429, American
  Institute of Physics Conference Series, ed. J.~L. {Hoffman}, J.~{Bjorkman},
  \& B.~{Whitney}, 244--247

\bibitem[{{Massey}(1997)}]{Massey:1997}
{Massey}, P. 1997, {A User's Guide to CCD Reductions with IRAF}
  (http://iraf.noao.edu/docs/recommend.html)

\bibitem[{{Naghizadeh-Khouei} \& {Clarke}(1993)}]{Naghizadeh-Khouei:1993}
{Naghizadeh-Khouei}, J., \& {Clarke}, D. 1993, \aap, 274, 968

\bibitem[{{Pereyra}(2000)}]{Pereyra:2000}
{Pereyra}, A. 2000, PhD thesis, Depto.~de Astronomia, Instituto Astron{\^o}mico
  e Geof{\'{\i}}sico, USP, Rua do Mat{\~a}o 1226 - Cidade Universit{\'a}ria
  05508-900 S{\~a}o Paulo SP - BRAZIL

\bibitem[{{Pereyra} \& {Magalh{\~a}es}(2004)}]{Pereyra:2004}
{Pereyra}, A., \& {Magalh{\~a}es}, A.~M. 2004, \apj, 603, 584

\bibitem[{{Serkowski}(1958)}]{Serkowski:1958}
{Serkowski}, K. 1958, \actaa, 8, 135

\bibitem[{{Serkowski} {et~al.}(1975){Serkowski}, {Mathewson}, \&
  {Ford}}]{Serkowski:1975}
{Serkowski}, K., {Mathewson}, D.~S., \& {Ford}, V.~L. 1975, \apj, 196, 261

\bibitem[{{Trujillo-Bueno} {et~al.}(2002){Trujillo-Bueno}, {Moreno-Insertis},
  \& {Sanchez Martinez}}]{Trujillo-Bueno:2002}
{Trujillo-Bueno}, J., {Moreno-Insertis}, F., \& {Sanchez Martinez}, F. 2002,
  {Astrophysical Spectropolarimetry} (Cambridge University Press)

\bibitem[{{Turnshek} {et~al.}(1990){Turnshek}, {Bohlin}, {Williamson}, {Lupie},
  {Koornneef}, \& {Morgan}}]{Turnshek:1990}
{Turnshek}, D.~A., {Bohlin}, R.~C., {Williamson}, II, R.~L., {et~al.} 1990,
  \aj, 99, 1243

\end{thebibliography}

\end{document}